\documentclass[twocolumn,reprint,aps,prd,superscriptaddress,nofootinbib]{revtex4-1} 
\usepackage[utf8]{inputenc}
\usepackage[english]{babel}
\usepackage[toc,page]{appendix}
\usepackage{amsmath}
\usepackage{mathrsfs}
\usepackage{amsfonts}
\usepackage{amssymb}
\usepackage{bbm}
\usepackage{graphicx}
\usepackage{xcolor}
\usepackage{hyperref}
\hypersetup{
    pdftitle={},
    colorlinks=true,
    linkcolor=blue,
    citecolor=blue,
    filecolor=blue,
    urlcolor=blue
}
\usepackage{cleveref}
\usepackage{feynmp}
\usepackage{feynmp-auto}

\newcommand{\xx}{\mathbf{x}}
\newcommand{\yy}{\mathbf{y}}
\newcommand{\EE}{\mathbf{E}}
\newcommand{\BB}{\mathbf{B}}
\newcommand{\Tr}{\mathrm{Tr}}
\newcommand{\dd}{\mathrm{d}}
\newcommand{\SU}{\mathrm{SU}}
\newcommand{\su}{\mathfrak{su}}
\newcommand{\ihat}{\hat{\mathbf{i}}}
\newcommand{\jhat}{\hat{\mathbf{j}}}
\newcommand{\khat}{\hat{\mathbf{k}}}
\newcommand{\pp}{\mathbf{p}}
\newcommand{\rr}{\mathbb{R}}
\newcommand{\zz}{\mathbb{Z}}
\newcommand{\dgm}{\mathrm{Dgm}}
\newcommand{\Pfrak}{\mathfrak{P}}

\begin{document}

\title{Probing universal dynamics with topological data analysis in a gluonic plasma} 
\author{Daniel Spitz}
\email{spitz@thphys.uni-heidelberg.de}
\affiliation{Institute for Theoretical Physics, Heidelberg University, Philosophenweg 16, 69120 Heidelberg, Germany}

\author{Kirill Boguslavski}
\affiliation{Institute for Theoretical Physics, Vienna University of Technology, Wiedner Hauptstraße 8-10/136, 1040 Vienna, Austria}

\author{Jürgen Berges}
\affiliation{Institute for Theoretical Physics, Heidelberg University, Philosophenweg 16, 69120 Heidelberg, Germany}

\begin{abstract}
We study nonequilibrium dynamics of SU(2) lattice gauge theory in Minkowski space-time in a classical-statistical regime, where characteristic gluon occupancies are much larger than unity. 
In this strongly correlated system far from equilibrium, the correlations of energy and topological densities show self-similar behavior related to a turbulent cascade towards higher momentum scales.
We employ persistent homology to infer topological features of the gluonic plasma via a hierarchy of simplicial and cubical complexes. 
All topological observables under investigation are also manifestly gauge invariant and are shown to exhibit self-similar evolution, which relate the spatial and temporal properties of the plasma in terms of universal scaling exponents and functions. 
The findings may help to understand the early stages of heavy-ion collisions in the limit of high energies, and our methods can also facilitate the topological analysis of other complex systems such as encountered in experiments with ultracold quantum gases. 
\end{abstract}

\maketitle

\section{Introduction}
Collective phenomena are ubiquitous during the thermalization of the quark-gluon plasma (QGP) generated in ultrarelativistic collisions of heavy nuclei at the Relativistic Heavy Ion Collider (RHIC) or the Large Hadron Collider (LHC)~\cite{Busza:2018rrf, Schlichting:2019abc, Berges:2020fwq}. 
Most of the freed quarks and gluons during the first yoctosecond of the collision originate from the small-$x$ regime, giving rise to the notion of gluon saturation and large gluon occupations~\cite{Gribov:1983ivg,Gelis:2010nm}. 
The `bottom-up' scenario provides a systematic description for the subsequent thermalization of the pre-equilibrium QGP~\cite{Baier:2000sb}. 
Its early stage is governed by gluon over-occupation at low momenta, such that the weakly coupled system can be accurately mapped onto classical-statistical Yang-Mills theory~\cite{Aarts:2001yn, Mueller:2002gd, Skullerud:2003ki, Jeon:2004dh, Arrizabalaga:2004iw, Kurkela:2012hp}, which sets the model for this study.
For simplicity we focus on homogeneous initial states in non-expanding Minkowski geometry, while more realistic extensions feature also longitudinal expansion~\cite{Bjorken:1982qr}. 

In this work we compute the nonequilibrium time evolution of an initially over-occupied gluonic plasma. 
For a real-time SU(2) lattice gauge theory in 3+1 dimensions in the classical-statistical regime, we analyze gauge-invariant observables using energy and topological densities and their fluctuations, which evade the typical drawbacks of using gauge-fixed field correlation functions. 
In order to infer non-local geometric information about the plasma's evolution, we use persistent homology~\cite{otter2017roadmap, chazal2021introduction, edelsbrunner2022computational}. 
Persistent homology computes topological structures that appear in a hierarchy of simplicial or cubical complexes constructed from the field data and features measures of their dominance (persistence). 
It behaves stable with respect to perturbations on the data, and can be efficiently calculated~\cite{otter2017roadmap, chazal2021introduction, edelsbrunner2022computational}. 
We devise gauge-invariant persistent homology observables and demonstrate that they exhibit self-similar evolution in terms of universal scaling exponents and functions.

Identifying universal phenomena in the time evolution of the QGP off equilibrium provides a rich line of understanding the otherwise complicated and theoretically challenging relaxation dynamics of strongly interacting nuclei~\cite{Berges:2020fwq}. 
For the gluonic plasma these include universal time dependences related to nonthermal fixed points~\cite{Berges:2013eia, Berges:2013fga, Berges:2014bba}, and the universal approach to local thermal equilibrium governed by viscous hydrodynamics~\cite{Heller:2015dha, Romatschke:2017vte, Soloviev:2021lhs}. 
Nonthermal fixed points provide far-from-equilibrium attractor solutions characterized by universal dynamical self-similarity. 
They have first been discussed for relativistic and non-relativistic scalar theories, partly in the context of reheating of the early universe after inflation~\cite{Micha:2002ey, Micha:2004bv, Berges:2008wm, Berges:2013lsa, PineiroOrioli:2015cpb}. 
In recent years, dynamical self-similarity has been found experimentally for ultracold Bose gases~\cite{Prufer:2018hto, Erne:2018gmz, Glidden:2020qmu}, which includes evidence for its universality across very different initial conditions.

For the gluonic plasma a dynamically separating hierarchy of momentum scales characterizes transport processes related to nonthermal fixed points. 
In the infrared, the string tension scale encodes the area dependence of spatial Wilson loops~\cite{Mace:2016svc, Berges:2017igc}, going along with condensation far from equilibrium~\cite{Berges:2019oun}. 
The Debye mass describing the electric screening scale evolves dynamically towards the infrared, too~\cite{Kurkela:2012hp, Mace:2016svc}, and the hard scale indicates energy transport towards the ultraviolet in a universal self-similar process~\cite{Berges:2008mr, Kurkela:2011ti, Kurkela:2012hp, Berges:2012ev, Schlichting:2012es, Berges:2013fga, AbraaoYork:2014hbk, Mace:2016svc, Boguslavski:2018beu, Boguslavski:2019fsb}.

Such universal behavior is typically probed using variants of occupation numbers.
These are constructed as two-point correlation functions of gluon fields, which are not gauge invariant and contain limited information on the plasma.
Therefore, it is instrumental to go beyond this description using gauge invariant observables.
Energy densities and their fluctuations are relevant quantities for the hydrodynamical description of the later fluid dynamics of the QGP in heavy ion collisions~\cite{Florkowski:2017olj, Romatschke:2017ejr}.
Topological densities are of importance for the structure of the vacuum of quantum chromodynamics (QCD), linked to chiral properties of quarks in the QGP via anomalies and of central relevance for topological effects such as sphaleron transitions~\cite{Mace:2016svc}.
For these reasons the study of two-point correlation functions of energy and topological (charge) densities is promising, which involve gauge invariant four-point correlation functions in the gluon fields. 

A complementary viewpoint on universal dynamics beyond local correlation functions can be based on persistent homology~\cite{otter2017roadmap, chazal2021introduction, edelsbrunner2022computational}. 
A first study of persistent homology dynamics in the vicinity of a nonthermal fixed point in scalar field theory demonstrated its ability to reveal self-similar behavior~\cite{Spitz:2020wej}. 
Persistent homology has been applied in the related context of critical phenomena~\cite{donato2016persistent, speidel2018topological,santos2019topological,olsthoorn2020finding,Sale:2021xsq, tran2021topological,cole2021quantitative,Sehayek:2022lxf}, notably confinement in non-Abelian gauge theory~\cite{Sale:2022qfn, Spitz:2022tul} and corresponding effective models~\cite{Hirakida:2018bkf, Kashiwa:2021ctc}. 

This paper is organized as follows. 
In \Cref{SecCorrelators} we provide details on the lattice setup and discuss energy and topological density correlations, which show clear manifestations of the direct cascade related to energy transport. 
We argue that dynamical scaling exponents can be understood from occupation numbers and the energy-momentum conservation Ward identity. 
We suggest an explanation for the accurate matching of energy and topological density correlations. In \Cref{SecPersHom} we introduce persistent homology, and discuss the self-similarity of Betti number distributions. 
We explain how this can heuristically be understood from the behavior of correlation functions in conjunction with the bounded packing of topological features in the constant lattice volume. 
Finally, in \Cref{SecConclusions} we conclude and provide an outlook. 
In the Appendixes we provide details on the lattice setup and occupation numbers, and discuss further theoretical aspects and more of our results for persistent homology.

\section{Direct cascade in energy and topological density correlators}\label{SecCorrelators}
Energy and topological density correlators show dynamical self-similarity indicative of an energy cascade. 
In this section we first describe the lattice simulations and the examined observables and proceed with an analysis of energy and topological density correlators. 

\subsection{Real-time non-Abelian gauge theory on the lattice}
We consider $\SU(N_c)$ gauge theory for $N_c=2$ on a cubic lattice with Minkowski metric, spatial extent $N_s^3$, and temporal and spatial lattice spacings $\dd t$ and $a_s$, respectively. 
Spatially periodic boundary conditions are applied. 
The volume of the spatial lattice $\Lambda_s$ is $V=a_s^3N_s^3$. 
We study the theory in temporal-axial gauge, $A_0\equiv 0$, and formulate it in terms of $\su(N_c)$-valued \mbox{(chromo-)}electric fields $E_i(t,\xx)$ and $\SU(N_c)$-valued link variables $U_i(t,\xx)$, $i=1,2,3$. 
The latter are related to the gauge fields $A_i(t,\xx)$ via $U_i(t,\xx)\approx \exp(i g a_s A_i(t,\xx))$ up to lattice corrections, with $g$ the gauge coupling. 

We study the system in the weakly coupled regime, $g\ll 1$, and consider highly occupied gluonic initial conditions. 
Then, at sufficiently early times the full quantum dynamics is accurately described by the classical-statistical approximation \cite{Aarts:2001yn, Mueller:2002gd, Skullerud:2003ki, Jeon:2004dh, Arrizabalaga:2004iw, Kurkela:2012hp,Berges:2013lsa}. 
Specifically, we sample over Gaussian initial conditions with variances
\begin{subequations}
\label{eq:init_cond}
\begin{align}
\langle AA\rangle (t{=}0,\pp) = &\; \frac{Q}{g^2 |\pp|^2}\, \theta(Q-|\pp|)\,,\\
\langle EE\rangle(t{=}0,\pp) = &\; \frac{Q}{g^2}\, \theta(Q-|\pp|)\,,
\end{align}
\end{subequations}
where averages over transverse polarizations and color degrees are implied. 
Here the momentum scale $Q$ determines the width of the initial momentum-space distribution.
In particular, these correlation functions are related to the distribution function of gluonic quasiparticles as $f(t{=}0,\pp) \simeq |\pp|\langle AA\rangle (t{=}0,\pp) \simeq \langle EE\rangle(t{=}0,\pp) / |\pp|$, see \Cref{AppendixOccupationNumbers}.
Such gluon over-occupation up to a gluon saturation scale is characteristic for a state shortly after the collision of heavy ions \cite{Berges:2020fwq}.
More details on how initial conditions are generated can be found in \cite{Boguslavski:2018beu, Berges:2013fga}.
The time evolution for the lattice fields $E_i(t,\xx)$ and $U_i(t,\xx)$ results from solving the classical equations of motion. 
These are the Hamilton equations for the instantaneous classical lattice Hamiltonian
\begin{align}\label{EqHamiltonian}
H(t)=&\; \frac{1}{2g^2}\sum_{\xx\in\Lambda_s}\bigg[ \Tr(\EE(t,\xx)^2)\nonumber\\
&\qquad + \frac{4}{a_s^4}\sum_{j>k}[N_c - \mathrm{Re}\,\Tr(U_{jk}(t,\xx))]\bigg]\\
=: &\; \sum_{\xx\in\Lambda_s} T^{00}(t,\xx)
\,,
\end{align}
with elementary spatial plaquette variables
\begin{equation}
U_{jk}(t,\xx) = U_j(t,\xx)U_k(t,\xx + \jhat) U_j^{\dagger}(t,\xx+ \khat) U_k^\dagger(t,\xx)\,,
\end{equation}
and $\jhat$, $\khat$ unit lattice vectors in directions $j$, $k$. 
We refer to $T^{00}(t,\xx)$ as the energy density, and emphasize its gauge-invariance.
Specifically, the discretized lattice equations of motion derived from the classical Hamiltonian \eqref{EqHamiltonian} read \cite{Boguslavski:2018beu}
\begin{subequations}\label{EqEOM}
\begin{align}
U_i(t+\dd t/2,\xx) =&\; e^{i\,\dd t \, a_s g E_i(t,\xx)} U_i(t-\dd t/2, \xx)\,,\label{EqEOMU}\\
g E_i(t+\dd t,\xx) =&\; g E_i(t,\xx) - \frac{\dd t}{a_s^3}\sum_{j\neq i}\bigg[U_{ij}(t-\dd t/2,\xx)\nonumber\\
&\qquad + U_{i(-j)}(t-\dd t/2,\xx)\bigg]_{\mathrm{ah}},\label{EqEOME}
\end{align}
\end{subequations}
where 
\begin{equation}
U_{i(-j)}(t,\xx)=U_i(t,\xx)U_j^{\dagger}(t,\xx+\ihat-\jhat)U^\dagger_i(\xx-\jhat)U_j(\xx-\jhat)\,.
\end{equation}
The anti-Hermitian part of a matrix $U$ is defined as
\begin{equation}
[U]_{\mathrm{ah}}=\frac{-i}{2}\bigg(U-U^\dagger - \frac{1}{N_c}\Tr(U-U^\dagger)\bigg)\,.
\end{equation}
Additionally, Gauss' law has to be satisfied. 
Once fixed initially, which is accomplished using a projection algorithm onto the constraint surface \cite{Moore:1996qs}, it is preserved by the time evolution of \Cref{EqEOM}.

Finally, observables are computed as the average over sampled initial conditions.
We set $N_s=512$, $\dd t / a_s = 0.05$ and $Qa_s = 0.125$, 
%, and explicitly verified approximate independence of our results from unphysical lattice parameters. 
and explicitly verified insensitivity of our results to variations of the lattice spacing, specifically upon comparison with results for $Q a_s = 0.25$ with the same lattice volume.%
\footnote{The later introduced Betti numbers are expected to scale proportional to system volume, based on mathematical theorems \cite{hiraoka2018limit,spitz2020self}. 
The properly rescaled quantities and scaling exponents are insensitive to volume changes, which we verified for $N_s=256$ compared to $N_s=512$ for the same lattice spacing, see \Cref{AppendixFiltrationParameterDependence}.}
Accordingly, no renormalization is applied to any of the computed observables. 
Results are shown for a single run, which for our large lattice agree with results averaged over multiple runs up to statistical fluctuations. 
Rescaling the fields $A_i\to gA_i$ and $E_i\to gE_i$, the equations of motion and initial conditions become independent of the coupling $g$.
This formally corresponds to the classical field limit $g \to 0$ while keeping $g^2 f$ fixed.
For more details on the lattice formulation we refer to \cite{Boguslavski:2018beu}. Based on \cite{Berges:2008zt, Ipp:2010uy, Berges:2017igc}, we expect similar results as in this work for $\SU(N_c)$ theories with $N_c \geq 2$. 

In conjunction with energy densities, we study the dynamics of topological densities $q(t,\xx)$, which are space-time integrands of Chern-Simons numbers. 
In the continuum, we have $q\sim \Tr(\EE\cdot \BB)$, with the {(chromo-)}magnetic field $\BB$. 
On the lattice we define a topological density as
\begin{equation}\label{EqTopologicalDensity}
q(t,\xx): = -\frac{1}{32\pi^2}\Tr (\EE^{\mathrm{av}}(t,\xx)\cdot \BB^{\mathrm{av}}(t,\xx))\,,
\end{equation}
where $\EE^{\mathrm{av}}(t,\xx)$ and $\BB^{\mathrm{av}}(t,\xx)$ are $\SU(N_c)$-valued clover-leaf variants of lattice electric and magnetic fields, averaging contributions of neighboring lattice sites as detailed in \Cref{AppendixLattice}. 
We emphasize that $q(t,\xx)$ defined through \Cref{EqTopologicalDensity} is gauge-invariant.

\subsection{Correlations of energy and topological densities}\label{SecCorrEnergyTopDensities}
Connected two-point correlation functions of the energy density $T^{00}(t,\xx)$ are constructed via
\begin{align}
C_{T^{00}}(t,\Delta \xx) :=&\; \sum_{\xx\in\Lambda_s} \langle T^{00}(t,\xx+\Delta\xx) T^{00}(t,\xx)\rangle_c\nonumber\\
=&\;\sum_{\xx\in\Lambda_s}\langle (T^{00}(t,\xx+\Delta \xx) - \bar{T}^{00})\nonumber\\
&\qquad\qquad\qquad\times (T^{00}(t,\xx)-\bar{T}^{00})\rangle
\end{align}
with the volume-averaged energy density
\begin{equation}
\bar{T}^{00} = \frac{1}{N_s^3}\sum_{\yy\in\Lambda_s} \langle T^{00}(t,\yy)\rangle\,,
\end{equation}
which remains constant in time in the simulations as required by total energy conservation. 
We Fourier-transform to lattice momentum space in $x$-direction,
\begin{equation}
C_{T^{00}}(t,\tilde{p}_x) = \sum_{\Delta x=0}^{N_s-1}C(t,\Delta\xx{=}(\Delta x,0,0))\,e^{-i\tilde{p}_x\Delta x}\,,
\end{equation}
with ${\tilde{p}_x \in 2\pi/(a_s N_s) \{-N_s/2,\dots,N_s/2-1\}}$ lattice momentum, corresponding to physical momentum ${p_x = (2/a_s) \sin(\tilde{p}_x a_s/2)}$. 
Finally, we set
\begin{equation}
\langle (T^{00}(t,p_x))^2\rangle_c := C_{T^{00}}(t,\tilde{p}_x)\,.
\end{equation}
In continuum this correlator corresponds to 
\begin{equation}
 \langle (T^{00}(t,p_x))^2\rangle_c = \int_{p_y,p_z} \langle T^{00}(t,\pp)(T^{00}(t,\pp))^*\rangle_c\,,
\end{equation}
where $T^{00}(t,\pp) = \int_{\xx} T^{00}(t,\xx)\exp(-i\pp \xx)$ with the notation $\int_\xx = \int_V \dd^3 x$ and $\int_{p_i} = \int \dd p_i/(2\pi)$.
The topological density two-point correlation function $\langle (q(t,p_x))^2\rangle_c$ is defined analogously.

\begin{figure*}
    \centering
	\includegraphics[scale = 0.8]{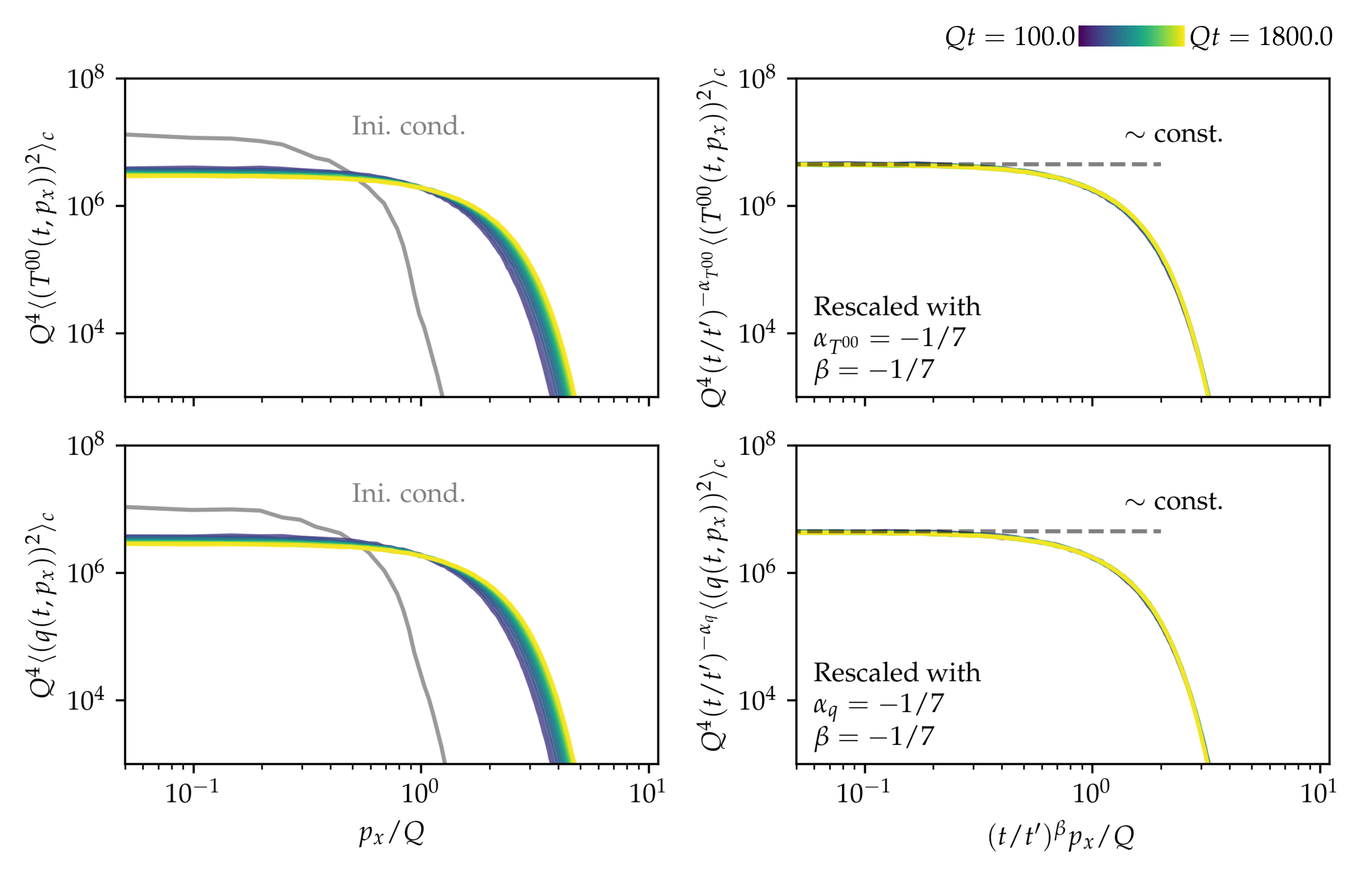}
	\caption{Connected two-point correlation functions of energy density $T^{00}(t,p_x)$ (top) and topological density $q(t,p_x)$ (bottom).
    Variants without (left) and with dynamical rescaling (right) are shown. Gray: Initial condition.}\label{FigT00qCorrels}
\end{figure*}

In \Cref{FigT00qCorrels} we show results for energy and topological density correlators, compared across times.
We find that they start off flat in momentum space and rapidly decay above the momentum scale $Q$. 
This indicates an equidistribution of low momentum fluctuations of energy and topological densities in momentum space.
After a short early-time dynamics (not displayed), the correlators remain constant in shape for times above $Qt=100$ and solely decrease in overall numbers while shifting to larger momenta. 
This is indicative of a nonthermal fixed point. 
We test for the characteristic dynamical self-similar scaling according to the scaling ansatz
\begin{equation}\label{EqT00ScalingAnsatz}
\langle (T^{00}(t,p_x))^2\rangle_c = (t/t')^{\alpha_{T^{00}}} \langle (T^{00}(t',(t/t')^{\beta}p_x))^2\rangle_c,
\end{equation}
for suitably chosen scaling exponents $\alpha_{T^{00}},\beta$ and a reference time $t'$ in the time interval when \Cref{EqT00ScalingAnsatz} describes the dynamics accurately. 
In the right column of \Cref{FigT00qCorrels} we display correlations rescaled according to the scaling ansatz (\ref{EqT00ScalingAnsatz}), where we suggestively set $\beta = -1/7$ and $\alpha_{T^{00}}=-1/7$. 
We find accurate matching of the rescaled correlations for the entire considered time range from $Qt=100$ to $Qt=1800$. 
The same observations apply to topological density correlations. 
In fact, both types of correlations agree with each other nearly exactly.
We argue for this at the end of this subsection, and primarily discuss $T^{00}$ correlations in the following, noting that similar arguments apply for topological density correlations.

One can heuristically motivate the scaling ansatz \eqref{EqT00ScalingAnsatz} by following a kinetic quasi-particle picture.
The Fourier-transformed local energy density fluctuations for sample $i$ can be approximated on an individual sample level as
\begin{equation}\label{EqT00OnePtScalingAnsatz}
T^{00}_i(t,\pp) \sim \omega_\pp \,f_i(t,\pp)\,.
\end{equation}
Here $\omega_\pp \simeq |\pp|$ is the gluon dispersion and $f_i(t,\pp)$ denotes a distribution function of gluon occupation numbers computed for a single classical-statistical sample.
It has previously been demonstrated that the gluon occupancy $\langle f_i(t,\pp)\rangle$ shows dynamical self-similarity for initial conditions similar to ours, scaling as \cite{Schlichting:2012es,Kurkela:2012hp, Berges:2013fga, AbraaoYork:2014hbk, Boguslavski:2018beu}
\begin{equation}
\label{EqScalingAnsatzOccupations}
\langle f_i(t,\pp)\rangle = (t/t')^{-4/7} \langle f_i(t',(t/t')^{-1/7}\pp)\rangle\,,
\end{equation}
even for single samples as for our simulations, see \Cref{AppendixOccupationNumbers}.
The scaling is reminiscent of a direct energy cascade to higher momenta since the momenta of hard modes grow as $(t/t')^{1/7} Q$ while the energy density remains constant during this process, $\langle \int_{\pp} T^{00}_i(t,\pp) \rangle = const$. 
Due to \Cref{EqT00OnePtScalingAnsatz} one expects that the exponent ${\beta=-1/7}$ governs the dynamics at hard momenta also for $\langle (T^{00}(t,p_x))^2\rangle_c$.

The scaling exponent $\alpha_{T^{00}}$ can be understood from the energy-momentum conservation Ward identity. 
We work in a continuum space-time with spatial volume $V$ in a general spatial dimension $d$. In \cite{Coriano:2017mux} via metric variations the identity
\begin{equation}
0 = \langle \partial_{x_1,\nu} T^{0\nu}(x_1)T^{00}(x_2)\rangle_c
\end{equation}
has been derived for correlations of the stress-energy tensor $T^{\mu\nu}(x)$, connected in the $T^{\mu\nu}(x_i)$. Integrating $x_1$ over $V$, we find using Gauss' theorem and periodic boundary conditions, such that no spatial boundary terms occur,
\begin{equation}
\label{eq:d0IntT00}
0 = \partial_{t_1}\int_\xx \langle T^{00}(t_1,\xx)T^{00}(t_2,\yy)\rangle_c\,.
\end{equation}
Expectation values in our simulations exhibit approximate homogeneity and isotropy in space, such that  
\begin{align}
\int_\xx \langle  T^{00}(t_1,\xx)T^{00}(t_2,\yy)\rangle_c =&\; \int_{\Delta \yy} \langle T^{00}(t_1,\mathbf{0})T^{00}(t_2,\Delta\yy)\rangle_c\nonumber\\
=&\; \int_{\Delta \xx} \langle T^{00}(t_1,\Delta\xx)T^{00}(t_2,\mathbf{0})\rangle_c\,,
\end{align}
where $\Delta \xx = \xx - \yy$ and $\Delta\yy = \yy-\xx$.
We note that
\begin{align}
&\partial_t \langle T^{00}(t,\xx)T^{00}(t,\yy)\rangle_c \nonumber\\
& = (\partial_{t_1} + \partial_{t_2})|_{t_1=t_2=t} \langle T^{00}(t_1,\xx)T^{00}(t_2,\yy)\rangle_c\,.
\end{align}
Analogous to \Cref{EqT00ScalingAnsatz} we assume a scaling ansatz in position space for $T^{00}$-correlations,
\begin{align}\label{EqT00ScalingAnsatzPositionSpace}
& \langle T^{00}(t,\xx)T^{00}(t,\yy)\rangle_c \nonumber\\
& = (t/t')^{\alpha'}\langle T^{00}(t',(t/t')^{-\beta}\xx)T^{00}(t',(t/t')^{-\beta}\yy)\rangle_c\,.
\end{align}
Together with classical-statistical simulations corresponding to symmetric operator ordering, denoted $\langle\cdot \rangle_s$, this yields
\begin{align}
0 = &\; (\partial_{t_1} + \partial_{t_2})|_{t_1=t_2=t}  \int_{\xx} \langle T^{00}(t_1,\xx)T^{00}(t_2,\yy)\rangle_{c,s}\nonumber\\
=&\; \partial_t \int_{\xx} \langle T^{00}(t,\xx) T^{00}(t,\yy)\rangle_{c,s}\nonumber\\
=&\, \partial_t \int_{\Delta\xx}\langle T^{00}(t,\Delta \xx) T^{00}(t,\mathbf{0})\rangle_{c,s}\nonumber\\
=&\; \partial_t \bigg[ (t/t')^{\alpha'+ d\beta} \int_{\Delta \xx'} \langle T^{00}(t',\Delta \xx')T^{00}(t',\mathbf{0})\rangle_{c,s}\bigg]\nonumber\\
= &\; \frac{\alpha' + d\beta}{t'}(t/t')^{\alpha'+ d\beta- 1}  \int_{\Delta \xx'} \langle T^{00}(t',\Delta \xx')T^{00}(t',\mathbf{0})\rangle_{c,s}\,,
\end{align}
with the substitution $\Delta \xx' = (t/t')^{-\beta}\Delta \xx$ and exploiting the empty boundary of the lattice torus due to periodic boundary conditions.
This is solved by $\alpha' = -d\beta$. 
We obtain the scaling behavior of $\langle (T^{00}(t,p_x))^2\rangle_c$ via momentum integration over directions $2,\dots,d$ and a Fourier transformation:
\begin{align}
&\langle (T^{00}(t,p_x))^2\rangle_c \nonumber\\
&= \int_{p_2,\dots,p_d} \int_{\Delta\xx}\int_{\xx} \langle T^{00}(t,\xx+\Delta\xx)T^{00}(t,\xx)\rangle_{c,s} e^{-i\pp\Delta\xx} \nonumber \\
&= (t/t')^\beta  \langle (T^{00}(t',(t/t')^\beta p_x))^2\rangle_c\,,
\end{align}
such that $\alpha_{T^{00}} = \beta$, which is consistent with \Cref{FigT00qCorrels}.

It remains to discuss the similarity of energy density $T^{00}\sim \Tr(\EE^2 +\BB^2)$ and topological density $q\sim \Tr(\EE\cdot \BB)$ correlations visible in \Cref{FigT00qCorrels} and giving rise to identical dynamical scaling behavior.
While their one-point functions are different, $\bar{T}^{00} = (0.106\pm 8.4\cdot 10^{-8})/Q^4$, $\bar{q} = (2.2\cdot 10^{-5}\pm 3.3\cdot 10^{-5})/Q^4$ (the error indicates fluctuations in time), their variances agree. 
This suggests that the colour and spatial directions of $\EE(t,\xx)$ and $\BB(t,\xx)$ are statistically independent and, by spatial homogeneity, identically distributed (i.i.d.). This is consistent with the observed ${\langle \Tr(\EE^2+ \BB^2) \rangle >0}$, ${\langle \Tr(\EE\cdot\BB)\rangle = 0}$, and identical higher-order connected correlation functions. 
Phrased differently, the space-time and colour components of the field strength tensor $F^{\mu\nu}(t,\xx)$ are i.i.d.\ up to antisymmetry of the indices.
Similar observations have been reported for two-point correlation functions of electric and magnetic fields \cite{Schlichting:2012es}. 
Note that this indicates suppressed contributions from non-trivial topological excitations such as sphalerons \cite{Mace:2016svc}.

The agreement of energy and topological densities provides further evidence for the universality of the nonthermal fixed point across different observables $\mathcal{O}$ with respect to its scaling exponent $\beta$. 
The amplitude exponent $\alpha_\mathcal{O}$ is linked to $\beta$ by constraints such as energy-momentum conservation. 
In the next section we provide a complementary study of this universality by analyzing non-local geometric observables instead of correlation functions.

\section{Direct cascade in persistent homology}\label{SecPersHom}
Persistent homology provides a quantitative means to extract topological structures from point clouds of data, and is based on the construction of a hierarchy of combinatorial objects like simplicial or cubical complexes. 
We will start this section with an intuitive introduction to the utilized alpha complexes and their persistent homology in \Cref{SecIntroAlpha}, followed by a discussion of dynamical self-similarity in persistent homology in \Cref{SecPersHomSelfSimilarity}. 
We then study the dynamics of topological structures in point clouds computed from energy and topological densities in \Cref{SecBettiResults}. 
We find clear indications for self-similar scaling associated to the energy cascade, which again reveals universality across observables. 
Moreover, we confirm a geometric relation between appearing exponents in persistent homology. 
In \Cref{SecPersistenceResults} we discuss persistence ratios of topological structures, i.e., spatially scale-invariant measures of their dominance, which form a set of dynamical invariants beyond self-similarity in the gluonic plasma.

Similarly to the findings for alpha complexes reported in this section, in \Cref{AppendixCubical} we discuss so-called cubical complexes, which give access to the persistent homology of level sets without spatial metric information entering the analysis, and are particularly suitable for pixelized data. 
In this approach the application of coarse-graining is inevitable to exclude lattice artefacts. 
Again, we find evidence for self-similarity associated to the energy cascade. 
Hence, the universal dynamics also encompasses the persistent homology of cubical complexes.
Since nonthermal fixed points are a scale-dependent phenomenon by means of self-similar transport across length scales, we focus on the results for alpha complexes here, which seem to reveal self-similar scaling for a wider time interval than cubical complexes.

\begin{figure}
    \centering
	\includegraphics[scale = 0.8]{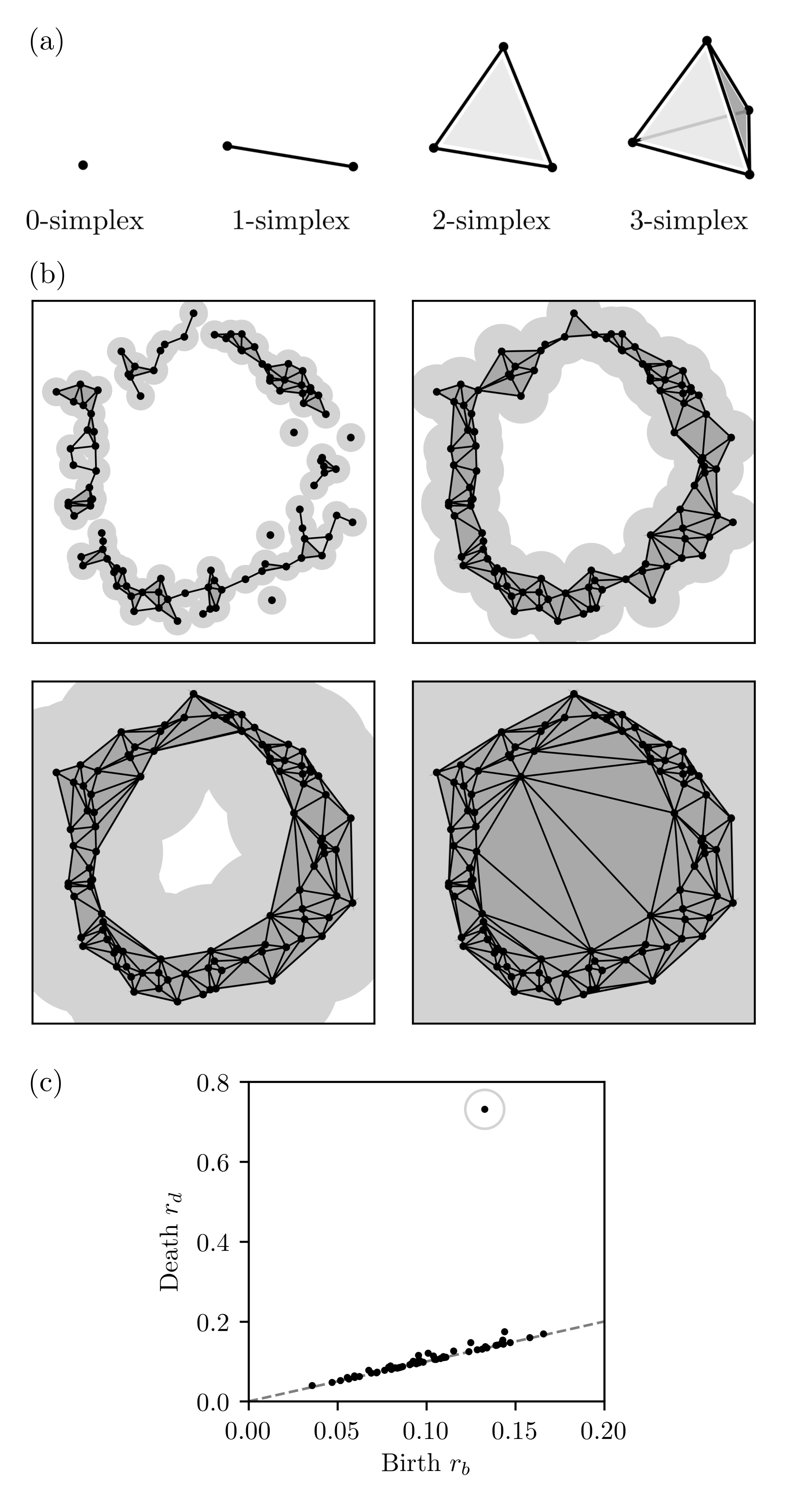}
	\caption{(a): Simplices of different dimensions; the 3-simplex is to be regarded as filled. (b): Alpha complexes of increasing radii. The point cloud in (b) consists of points sampled randomly from a circle with Gaussian noise added on their positions. (c): Persistence diagram of dimension-1 holes of the point cloud. Dashed line: the diagonal $r_b=r_d$.}\label{FigAlphaComplexes}
\end{figure}

\subsection{Introducing alpha complexes and persistent homology}\label{SecIntroAlpha}
Let $X\subset \rr^3$ be a point cloud, i.e., a finite set of distinct points.
\footnote{We assume the point cloud to be in general position, i.e., that no three points are collinear, that no four points lie on a single circle, and that no five points lie on a sphere.}
Alpha complexes of $X$ are computationally useful geometric simplicial complexes describing $X$. 
Their topology can be extracted algorithmically, which is facilitated by their combinatorial properties. 
Here we focus on an intuitive description, and in \Cref{AppendixMathPersHom} we deliver more mathematical details related to persistent homology. 
For a thorough introduction to topological data analysis we refer to the literature \cite{otter2017roadmap, chazal2021introduction, edelsbrunner2022computational}.

Simplices of different dimensions are illustrated in \Cref{FigAlphaComplexes}(a). 
We refer to a point as a 0-simplex, a line as a 1-simplex, a triangle as a 2-simplex, and a filled tetrahedron as a 3-simplex. 
A simplicial complex is a collection of simplices closed under taking boundaries. We construct alpha complexes of $X$. 
Trivially, we can connect two points in $X$ by a straight line, i.e., a 1-simplex. 
Similarly, through any three points in $X$ we can draw a unique circle and through any four points in $X$ we can draw a unique sphere. 
We call a circle or sphere empty if all points of $X$ lie on or outside of it. 
We can associate to any two, three and four points a 1-, 2- or 3-simplex connecting them, and a radius parameter. 
The latter can be defined as half of the length of the line between the two points, as the radius of the unique circle through the three points, or as the radius of the unique sphere through the four points, respectively. 
We finally define the \emph{alpha complex} $\alpha_r(X)$ of $X$ with radius $r$ to consist of empty simplices of arbitrary dimension $\leq 3$ with radius $\leq r$. 
The construction works analogously in an arbitrary number of dimensions. Alpha complexes are simplicial complexes.

In \Cref{FigAlphaComplexes}(b) we give two-dimensional examples of alpha complexes of increasing radii (indicated by gray disks around the points) for a cloud consisting of points sampled randomly from a circle with Gaussian noise added to their positions. 
We note that simplices which enter at larger radii have visually larger area.
For a finite point cloud there are finitely many different alpha complexes $\alpha_{r_i}(X)$ with $r_i< r_j$ for $i<j$, $i,j=1,\dots,K$ for an integer $K$. 
They form a \emph{filtration}:
\begin{equation}
\alpha_0(X)= \emptyset\subsetneq \alpha_1(X)\subsetneq \cdots \subsetneq \alpha_K(X) = \mathrm{Del}(X)\,.
\end{equation}
The final, `full' alpha complex $\mathrm{Del}(X)$ is also known as the Delaunay complex or Delaunay triangulation. 
That alpha complexes of increasing radii are nested can be seen in \Cref{FigAlphaComplexes}(b).

As we sweep through the filtration of alpha complexes $\alpha_r(X)$, topology changes appear. 
We observe in \Cref{FigAlphaComplexes}(b) that for the lowest radius the alpha complex consists of many distinct connected components. 
Upon increasing the radius the connected components successively merge with each other, such that the large loop gets \emph{born} in the alpha complex. 
The large loop \emph{persists} for a comparably large range of radii, until it \emph{dies} when filled entirely by triangles. 
Its topology does not `see' its thickening upon increasing the radius. 
We describe the topology of $\alpha_r(X)$ for a specific radius $r$ by homology, which can be explicitly computed. 
Homology can describe holes of different dimensions in complexes. 
A dimension-0 hole is a connected component, a \mbox{dimension-1} hole is an unfilled, planar-like loop, and a dimension-2 hole is an empty void in the complex.

\emph{Persistent homology} describes the changing topology of the entire filtration of alpha complexes at once. 
It associates to a dimension-$k$ hole a birth radius $r_b$ and a death radius $r_d> r_b$, such that the hole is present in the alpha complexes for all radii $r\in [r_b,r_d)$, with $r_b$ minimal and $r_d$ maximal. 
We can describe the persistent homology of the filtration of alpha complexes $\{\alpha_r(X)\}_r$ by the collection of all birth-death pairs $(r_b,r_d)$, which we call the \emph{persistence diagram}, denoted $\dgm_k(X)$. 
In \Cref{FigAlphaComplexes}(c) we display the dimension-1 persistence diagram of the noisy, approximately circular point cloud. 
We note that many features appear near the diagonal $r_b=r_d$, i.e., have persistence ratio $r_d/r_b$ near 1, represented by the tiny, short-lived unfilled loops which appear at smaller radii in \Cref{FigAlphaComplexes}(b). 
One prominent feature appears at large $r_d$, highlighted in \Cref{FigAlphaComplexes}(c). 
Having a high persistence ratio $r_d/r_b\simeq 5$, this feature is represented by the large circular structure present in the point cloud. 
In summary, persistent homology provides a quantitative means to extract topological structures from point cloud data, and incorporates persistence measures of their dominance.

Persistent homology has a number of useful theoretical properties. It is stable, i.e., perturbations of the input point cloud result in only slight changes of corresponding persistence diagrams with respect to suitable metrics \cite{cohen2007stability, cohen2010lipschitz}. 
Furthermore, most observables computed with persistent homology behave well with regard to statistical limits, and also large volume asymptotics exist, including notions of ergodicity \cite{hiraoka2018limit, spitz2020self}. 
The efficient computation of persistent homology is facilitated by a variety of suitable libraries for different programming languages, see \cite{otter2017roadmap} for an overview. 
We employ the versatile GUDHI library \cite{10.1007/978-3-662-44199-2_28}, and compute periodic alpha complexes to take the spatially periodic boundary conditions of our lattice into account.

\subsection{Self-similarity in persistent homology}\label{SecPersHomSelfSimilarity}
How can dynamical self-similarity manifest itself in persistent homology? Let $X(t)\subset \rr^d$ be a family of time-dependent point clouds. 
In \cite{Spitz:2020wej} the dimension-$k$ persistence pair distribution has been introduced as
\begin{equation}
\Pfrak_k(t,r_b,r_d) = \sum_{(r_b',r_d')\in \dgm_k(X(t))} \delta(r_b - r_b')\,\delta(r_d-r_d')\,.
\end{equation}
In general, its expectation value $\langle \Pfrak_k\rangle (t,r_b,r_d)$ exists and is not a sum of Dirac $\delta$-functions anymore \cite{chazal2018density, spitz2020self}. 
We note that $\langle\Pfrak_k\rangle(t,r_b,r_d)$ has support only for $r_d>r_b$.
We say that the expected persistence pair distribution $\langle \Pfrak_k\rangle (t,r_b,r_d)$ scales self-similarly in time if
\begin{align}
\label{EqPersPairDistribScalingAnsatz}
&\langle \Pfrak_k\rangle (t,r_b,r_d) \nonumber\\
&\quad = (t/t')^{-\eta_2} \langle \Pfrak_k\rangle (t',(t/t')^{-\eta_1} r_b,(t/t')^{-\eta_1}r_d)\,,
\end{align}
for suitable scaling exponents $\eta_1,\eta_2$ and arbitrary reference time $t'$ in the self-similar regime. 
This scaling ansatz is similar to the position space scaling ansatz for the $T^{00}$-correlator given in \Cref{EqT00ScalingAnsatzPositionSpace}.

\begin{figure*}[t]
    \centering
	\includegraphics[scale = 0.77]{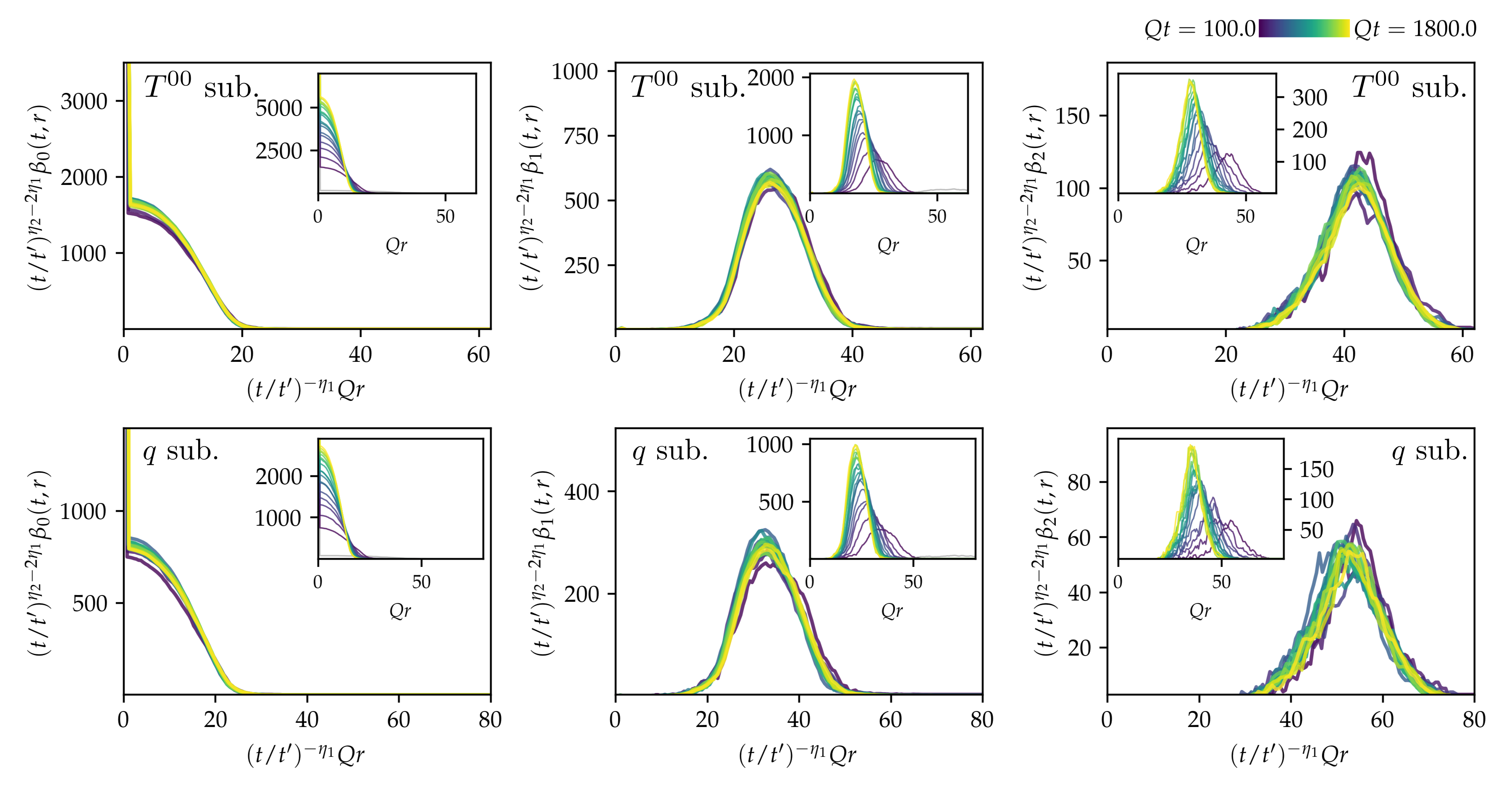}
	\caption{Betti number distributions of the alpha complex filtration for $T^{00}(t,\xx)$ (top) and $q(t,\xx)$ sublevel sets (bottom), with limiting values $\nu_{T^{00}} = -0.081/Q^4$ and $\nu_{q} = -0.13/Q^4$. Scaling exponents are set to $\eta_1 = -1/7$ and $\eta_2 = 5\eta_1 = -5/7$ in accordance with the packing relation. Insets show figures without rescaling. Distributions at initial time are displayed in gray.}\label{FigBettiAlpha}
\end{figure*}

From $\langle\Pfrak_k\rangle$ we can compute many persistent homology descriptors. 
For instance, the expected number of dimension-$k$ holes at radius $r$, known as the \emph{Betti number}, can be computed as
\begin{equation}
\beta_k (t,r) = \int_0^r \dd r_b \int_r^\infty \dd r_d \, \langle \Pfrak_k\rangle(t,r_b,r_d)\,.
\end{equation}
With \Cref{EqPersPairDistribScalingAnsatz} Betti numbers scale as
\begin{equation}\label{EqBettiScaling}
\beta_k(t,r) = (t/t')^{2\eta_1 - \eta_2}\beta_k(t',(t/t')^{-\eta_1}r)\,.
\end{equation}
$\langle\Pfrak_k\rangle$ also yields the total number of homology classes that appear in the filtration at arbitrary radius as
\begin{equation}
n_k(t) = \int_0^\infty \dd r_b \int_0^\infty \dd r_d \, \langle\Pfrak_k\rangle(t,r_b,r_d)\,,
\end{equation}
which scales as
\begin{equation}\label{EqNPersHomClassesScaling}
n_k(t) = (t/t')^{2\eta_1 - \eta_2} n_k(t')\,.
\end{equation}
The exponent $\eta_2$ thus encodes the power-law decline of the number of persistent homology classes for given birth and death radii $r_b,r_d$. 
Similarly, we can compute from $\langle\Pfrak_k\rangle$ the average death radius of dimension-$k$ holes as
\begin{align}
\langle r_{d,k} \rangle(t) =&\; \frac{1}{n_k(t)} \int_0^\infty \dd r_b \int_0^\infty \dd r_d\, r_d\, \langle\Pfrak_k\rangle(t,r_b,r_d)\nonumber\\
 =&\; (t/t')^{\eta_1} \langle r_{d,k}\rangle(t')\,.
\end{align}
The exponent $\eta_1$ describes the dynamical power-law blow-up of length scales associated to persistent homology.
For later use we note that the normalized distribution of persistence ratios $\pi = r_d/r_b$ can be computed from $\langle\Pfrak_k\rangle$ as
\begin{equation}
\Pi_k(t,\pi) = \frac{1}{n_k(t)}\int_0^\infty \dd r_b \, r_b\,\langle\Pfrak_k\rangle(t,r_b,\pi r_b)\,.
\end{equation}
For self-similar time evolutions all $\Pi_k(t,\pi)$ constitute invariants of motion. 
It has been conjectured in \cite{bobrowski2022universality} that the distributions $\Pi_k(t,\pi)$ are in fact universal across different processes to generate i.i.d.\ point clouds, thus invariant beyond self-similar time evolutions. 
We will confirm this in \Cref{SecPersistenceResults}.

To study the self-similar behavior of integrated quantities such as the total number of homology classes $n_k(t)$ in lattice simulations, persistent homology classes of sizes near the lattice spacing $a_s$ and near the lattice size $a_s N_s$ need to be explicitly excluded from the integrations. 
The employed procedure for persistence ratio distributions is described in \Cref{SecPersistenceResults}.

\subsection{Dynamics of homology classes in Betti numbers}\label{SecBettiResults}
Alpha complexes and their persistent homology require point clouds as input. 
Motivated by the findings of \Cref{SecCorrEnergyTopDensities}, we examine sublevel sets of energy and topological densities, constructed for $T^{00}_i(t,\xx)$ and $q_i(t,\xx)$, and computed from an individual classical-statistical sample $i$ as
\begin{subequations}
\begin{align}
X_{T^{00},\nu_{T^{00}},i}(t):=&\;\{\xx \in \Lambda_s\,|\, T^{00}_i(t,\xx)-\bar{T}^{00}\leq \nu_{T^{00}}\}\,,\\
X_{q,\nu_{q},i}(t):=&\; \{\xx\in\Lambda_s\,|\, q_i(t,\xx)\leq \nu_q\}\,.
\end{align}
\end{subequations}
For each such point cloud at time $t$ we compute the persistent homology of its filtration of alpha complexes. 
In the classical-statistical approximation we approximate expectation values of persistent homology descriptors as ensemble averages of observables computed for individual samples \cite{Spitz:2020wej}.

In \Cref{FigBettiAlpha} we show results for Betti number distributions of energy and topological density sublevel sets. 
We set $\nu_{T^{00}} = -0.081/Q^4$ and $\nu_q = -0.13/Q^4$, such that the point clouds comprise $20,000-50,000$ points which reflect the dynamics of local minima. 
For comparison, $\bar{T}^{00}$ and $\bar{q}$ have been given at the end of \Cref{SecCorrEnergyTopDensities}. 
We notice that dimension-0 Betti numbers monotonously decrease with increasing radii, which originates from connected components merging and thus decreasing in number.
Dimension-1 and -2 Betti numbers show distinct peaks. 
We observe peaks appearing at larger radii the larger the dimension of the holes is. 
Dimension-0 holes are present at lowest radii, required to merge with each other to form dimension-1 holes, which in turn need to die (getting filled with triangles) to form dimension-2 holes, and thus giving rise to the hierarchy. 
Focusing on the inset figures for $T^{00}_i$ and $q_i$ sublevel sets, we notice a continuous motion of holes to lower radii with time, across all dimensions. 
The numbers of holes increase, while the shapes of Betti number distributions qualitatively remain invariant. 
This suggests that topological structures in the alpha complexes decrease in size in the course of time.

The main figures are rescaled according to \Cref{EqBettiScaling}. 
We propose the exponents ${\eta_1 = -1/7}$, $\eta_2 = -5/7$. 
For these we observe that the rescaled distributions coincide to good accuracy, which is indicative of the validity of the self-similar scaling ansatz for persistent homology in \Cref{EqPersPairDistribScalingAnsatz}. 
The relation $\eta_2 = 5\eta_1$ is an example of the general packing relation first observed in \cite{Spitz:2020wej} and rigorously proven in \cite{spitz2020self}, and originates from the geometric bound that the constant volume puts on the number of holes. 
If holes decrease in size, more of them fit on the lattice.

We explicitly checked for the independence of the scaling behavior and qualitative shapes of Betti number distributions from the choice of $\nu_{T^{00}}$ and $\nu_q$ in nearby regimes, see \Cref{AppendixFiltrationParameterDependence}. 
We verified an approximate insensitivity from infrared and ultraviolet lattice cutoffs, such that the non-local features that persistent homology captures are well-resolved and sufficiently dense.

We conclude that the persistent homology of sublevel sets of energy and topological densities reveals self-similar scaling related to the energy cascade with the same spatial scaling exponent $\eta_1=\beta$ as for correlation functions. 
Yet, while energy densities are bounded by positivity, $T^{00}_i \geq 0$, this is not the case for topological densities. 
This generally results in different functional shapes through the involved point cloud construction. 
This is to be contrasted with correlation functions. 
Still, with regard to scaling exponents, we find that the universality of the self-similar dynamics related to the energy cascade encompasses persistent homology observables.

\begin{figure}
    \centering
	\includegraphics[scale = 0.77]{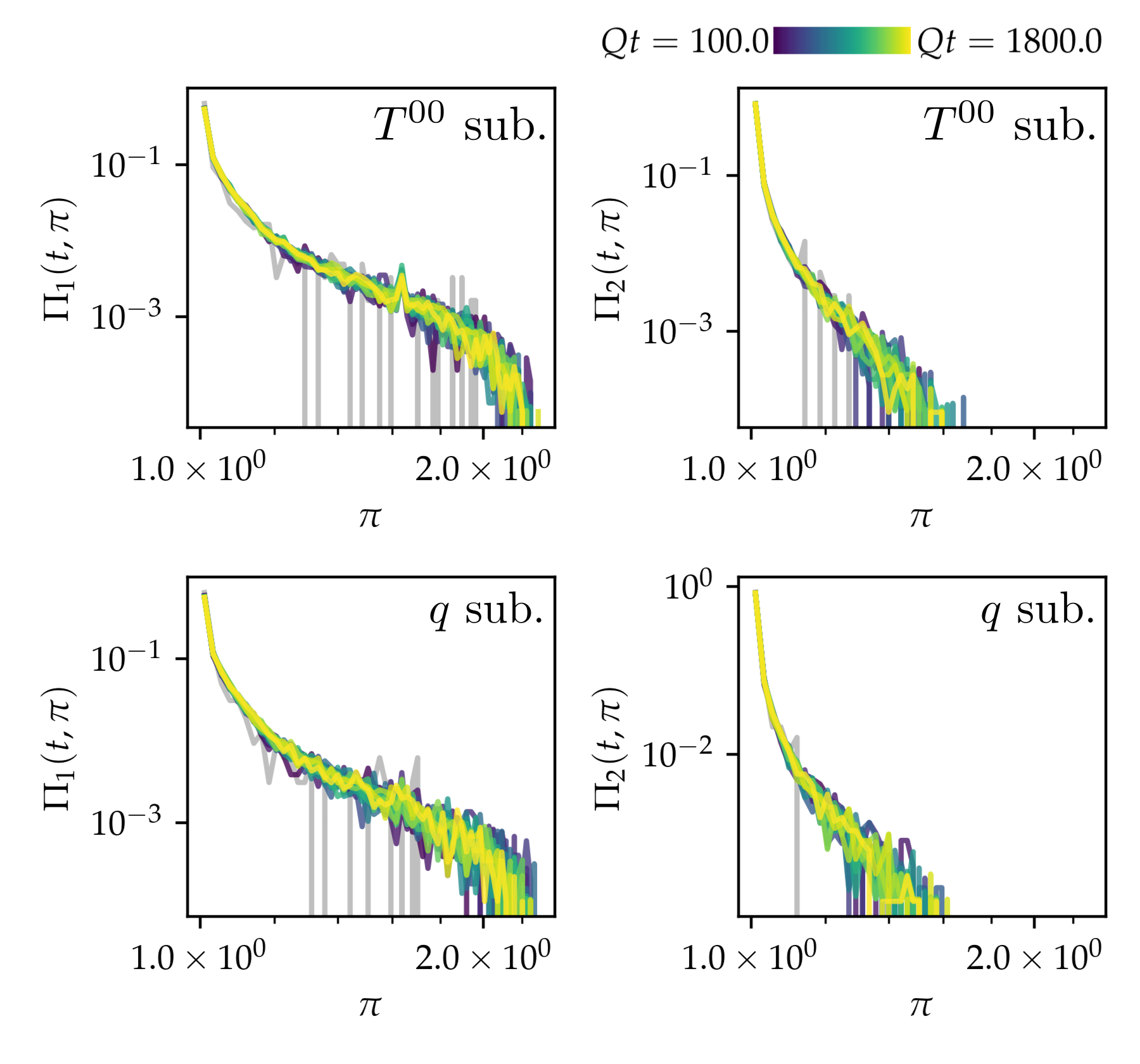}
	\caption{Normalized persistence ratio distributions of the alpha complex filtration for $T^{00}(t,\xx)$ (top) and $q(t,\xx)$ sublevel sets (bottom), with $\nu_{T^{00}} = -0.081/Q^4$ and $\nu_{q} = -0.13/Q^4$. Distributions at the initial time are displayed in gray.}
    \label{FigPersistencesAlpha}
\end{figure}

\subsection{Persistence ratio distributions}\label{SecPersistenceResults}
In \Cref{FigPersistencesAlpha} we display distributions of persistence ratios, $\Pi_k(t,\pi)$, for the energy and topological density sublevel sets $X_{T^{00},\nu_{T^{00}},i}(t)$ and $X_{q,\nu_q,i}(t)$. 
The distributions $\Pi_k(t,\pi)$ are particularly sensitive to lattice artefacts.
We explicitly removed the two most prominent lattice artefacts from the computed distributions by excluding in dimension 1 holes with $\pi=\sqrt{2}$ and $\pi=\sqrt{4/3}$, and in dimension 2 holes with $\pi=\sqrt{3/2}$ and $\pi=\sqrt{9/8}$. 
Square roots of quotients of small integers are unique to specific point configurations on the lattice originating from vertices of one up to few elementary lattice voxels. 
We checked that further lattice artifacts in this sense do not contribute significantly to the distributions shown in \Cref{FigPersistencesAlpha}.

The distributions $\Pi_k(t,\pi)$ decrease fast and have support roughly up to $\pi\approx 2$ in dimension 1 and up to $\pi\approx 1.6$ in dimension 2. 
Crucially, the distributions remain constant in time up to statistical fluctuations. 
Although showing more fluctuations at smaller $\Pi_k(t,\pi)$-values, this also applies to the initial time (indicated in gray). 
Thus, the time translation invariance of $\Pi_k(t,\pi)$ is not due to the specifics of a self-similar time evolution. 
The persistence ratio distributions $\Pi_k(t,\pi)$ rather constitute approximate invariants of motion for all $\pi$.

As already mentioned in \Cref{SecPersHomSelfSimilarity}, it has been conjectured that persistence ratio distributions are universal across i.i.d.\ point cloud generation processes \cite{bobrowski2022universality}. 
We regard the constancy in time of $\Pi_k(t,\pi)$ in \Cref{FigPersistencesAlpha} as a manifestation of this. 
An even stronger universality conjecture for a distribution of transformed $\pi$-values has been provided in the same work \cite{bobrowski2022universality}. 
This applies to our data, too, as we detail in \Cref{AppendixStrongUniversality}.

\section{Conclusions}\label{SecConclusions}
In this work we have studied energy and topological densities in real-time SU(2) lattice gauge theory simulations in the classical-statistical regime with over-occupied initial conditions. 
We found that energy and topological density correlation functions reveal self-similarity related to a direct energy cascade. 
Corresponding scaling exponents can be understood from a kinetic quasi-particle picture ($\beta$) and the energy-momentum conservation Ward identity ($\alpha(\beta)$). 

The approximate agreement of energy and topological density correlations hints at independently and identically distributed colour and space directions of $\EE(t,\xx)$ and $\BB(t,\xx)$.
Differences between energy and topological density correlations could have been due to topological excitations such as sphalerons. 
While topological excitations typically manifest themselves in coarsening dynamics due to their mutual annihilation, in our study both energy and topological density correlations instead showed a dynamical refinement of associated structures. 
This suggests that topological excitations are barely of relevance for the dynamics in gluonic plasmas at the length scales probed by the energy and topological densities for the examined over-occupied initial conditions.

In order to look for self-similar dynamics beyond local correlation functions, we constructed non-local observables based on persistent homology.
Specifically, we examined the filtration of alpha complexes of energy and topological density sublevel sets, and the filtration of cubical complexes of energy and topological densities (in \Cref{AppendixCubical}). 
Even for topological observables like Betti number distributions we observed self-similarity. 
The associated scaling exponents can be understood from the correlation functions ($\eta_1  =\beta$) and the packing relation ($\eta_2(\eta_1)$), which follows from the bounded number of topological features for a given non-expanding lattice volume. 
We emphasize that all observables investigated are gauge invariant, and the universality related to the direct energy cascade encompasses local and non-local observables. 
In particular, we have demonstrated in \Cref{AppendixCubical} that for cubical complex filtrations of sub- or superlevel sets where the spatial metric information does not enter, the packing relation applies as soon as there is \emph{any} bound on the filtration. 
This goes beyond previous findings for alpha complexes \cite{Spitz:2020wej}, and is of relevance to any persistent homology study of critical phenomena in or out of equilibrium.

More generally, persistent homology can provide a sensitive probe for non-local structures in lattice field data, as our results and persistent homology studies of the partly intricate phase structures of diverse physical models with non-local order parameters have been demonstrating \cite{olsthoorn2020finding,Sale:2021xsq, tran2021topological,cole2021quantitative,Sehayek:2022lxf,Sale:2022qfn, Spitz:2022tul}. 
Non-local excitations including topological defects can be important for a wide range of dynamical phenomena ranging from heavy-ion collisions to ultracold quantum gases \cite{Berges:2019oun,Lannig:2023fzf}. 
Beyond that, higher-order correlation functions are numerically hard to access but can contain important information in strongly correlated systems. 
Persistent homology can thus provide valuable observables in statistical contexts sensitive to higher-order and non-local correlations, in line with the results of the present study.

While we concentrated on gluonic plasma dynamics, understanding far-from-equilibrium universality classes is of relevance for a wide range of applications in complex many-body systems such as experiments with ultracold quantum gases.
Our study opens a pathway for their characterization including topological and non-local quantifiers.

\begin{acknowledgments}
We thank J.~Pawlowski, T.~Preis, J.~Urban and A.~Wienhard for discussions and collaborations on related topics. 
We acknowledge support by the Interdisciplinary Center for Scientific Computing (IWR) at Heidelberg University, where part of the numerical work has been carried out. 
This work is part of and supported by the Collaborative Research Centre, Project-ID No. 273811115, SFB 1225 ISOQUANT of the Deutsche Forschungsgemeinschaft (DFG, German Research Foundation), and supported by the  DFG under Germany's Excellence Strategy EXC 2181/1 - 390900948 (the Heidelberg STRUCTURES Excellence Cluster). 
This project was funded in part by the Austrian Science Fund (FWF) under project P 34455-N.
\end{acknowledgments}

\vspace{0.5cm}

\appendix

\begin{widetext}
%\section{Details of the lattice setup}\label{AppendixLattice}
\section{Clover-leaf electric and magnetic fields}\label{AppendixLattice}
This appendix is devoted to the utilized clover-leaf electric and magnetic fields.
%This appendix is devoted to details on the lattice implementation. 
In temporal-axial gauge the clover-leaf electric field $\EE^{\textrm{av}}$ is defined from the electric fields $E_i(t,\xx)$ as the Lie algebra element
\begin{align}
E^{\textrm{av}}_i(t,\xx):=&\; \frac{1}{4} \left[E_i(t+\dd t/2,\xx) + U^\dagger_{i}(t,\xx-\hat{i})E_i (t+\dd t/2,\xx-\hat{i})U_{i}(t,\xx-\hat{i}) \right.\nonumber\\
&\qquad  \left. + E_i(t-\dd t/2,\xx) + U^\dagger_{i}(t-\dd t,\xx-\hat{i})E_i(t-\dd t/2,\xx-\hat{i})U_{i}(t-\dd t,\xx-\hat{i})\right]_{\mathrm{ah}}.
\label{EqEavDef}
\end{align}
Numerically, due to small deviations $\mathcal{O}(\dd t)$ compared to the full definition (\ref{EqEavDef}), we employ the following electric field variant
\begin{equation}
E^{\textrm{av}}_i(t,\xx) = \frac{1}{2} \left[E_i(t+\dd t/2,\xx)+ U^\dagger_{i}(t,\xx-\hat{i})E_i (t+\dd t/2,\xx-\hat{i})U_{i}(t,\xx-\hat{i})\right]_{\mathrm{ah}},\label{EqEavDefNumerical}
\end{equation}
which transforms under a gauge transformation $V$ as $E_i^{\mathrm{av}}(t,\xx)\mapsto V(t,\xx)E_i^{\mathrm{av}}(t,\xx)V^\dagger(t,\xx)$.
We define a clover-leaf magnetic field as the Lie algebra element
\begin{align}
B^{\mathrm{av},i}(t,\xx) =&\; \frac{1}{4}\epsilon^{ijk} \left[U_{jk}(t,\xx)  + U_j^\dagger(t,\xx-\hat{j})U_{jk}(t,\xx-\hat{j}) U_j(t,\xx-\hat{j})\right. +  U_k^\dagger(t,\xx-\hat{k})U_{jk}(t,\xx-\hat{k})U_k(t,\xx-\hat{k})  \nonumber\\
&\qquad \left. +U_k^\dagger(t,\xx-\hat{k})U_j^\dagger(t,\xx-\hat{j}-\hat{k})U_{jk}(t,\xx-\hat{j}-\hat{k})U_j(t,\xx-\hat{j}-\hat{k})U_k(t,\xx-\hat{k})\right]_{\mathrm{ah}},\label{EqBavDef}
\end{align}
\end{widetext}
which also transforms gauge covariantly under a gauge transformation $V$ as $B^{\mathrm{av},i}(t,\xx)\mapsto V(t,\xx)B^{\mathrm{av},i}(t,\xx)V^\dagger(t,\xx)$.

As usual in lattice studies, a spatially symmetrized (thus clover-leaf) version of the topological density is used \cite{DiVecchia:1981aev,Rothe:1992nt}, defined as
\begin{equation}\label{EqDefEuclideanTopDensity}
q(x) = -\frac{1}{2^9 \pi^2}\sum_{\mu\nu\rho\sigma=\pm 0}^{\pm 3} \tilde{\epsilon}^{\mu\nu\rho\sigma} \Tr(U_{\mu\nu}(x)U_{\rho\sigma}(x)).
\end{equation}
Here, the fully antisymmetric $\tilde{\epsilon}^{\mu\nu\rho\sigma}=\tilde{\epsilon}_{\mu\nu\rho\sigma}$ is defined through $1 = \tilde{\epsilon}_{0123} = -\tilde{\epsilon}_{1023} = -\tilde{\epsilon}_{(-0)123}$. 
In terms of the previously defined clover-leaf electric and magnetic fields, $q(t,\xx)$ can be written in temporal-axial gauge as
\begin{equation}\label{EqTopDensityDef}
q(t,\xx) = - \frac{1}{32\pi^2} \sum_{i=1}^3 \Tr ( E^{\mathrm{av}}_i(t,\xx)B^{\mathrm{av},i}(t,\xx)).
\end{equation}
This expression coincides with \cite{Moore:1996wn}, is gauge-invariant and symmetric under parity transformations due to the employed clover-leaf electric and magnetic fields.

\section{Self-similarity in occupation numbers}\label{AppendixOccupationNumbers}
One typically investigates dynamical scaling behavior in the vicinity of nonthermal fixed points using occupation numbers. 
In a non-Abelian gauge theory, the definition of occupation numbers suffers from ambiguities. 
We focus on occupation numbers defined from electric fields as \cite{Boguslavski:2018beu}
\begin{subequations}\label{EqfEE_defs}
\begin{align}
& f_{EE}(t,p) \nonumber\\
&\quad =\; \frac{1}{(N_c^2-1) V p} P_T^{ij}(\pp) \langle E^a_i(t,\pp) (E_j^a(t,\pp))^*\rangle,\\
&f_{\dot{E}\dot{E}}(t,p)\nonumber\\
 &\quad =\; \frac{1}{(N_c^2-1) V p^3} P_T^{ij}(\pp) \langle (\partial_t E^a_i(t,\pp)) (\partial_t E_j^a(t,\pp))^*\rangle,
\end{align}
\end{subequations}
with color decomposition $E_j(t,\pp) = E_j^a(t,\pp) T^a$ of the Fourier-transformed electric field, $T^a$ generators of $\su(N_c)$, $p \equiv |\pp|$, and the transversal polarization projector $P^{ij}_T(\pp)$ defined as in \cite{Boguslavski:2018beu}. 
For simplicity we ignore effects from hard thermal loop self-energies, in particular screening masses, since we focus on hard momentum modes while they would alter correlators primarily at soft momenta.

\begin{figure}
    \centering
	\includegraphics[scale = 0.77]{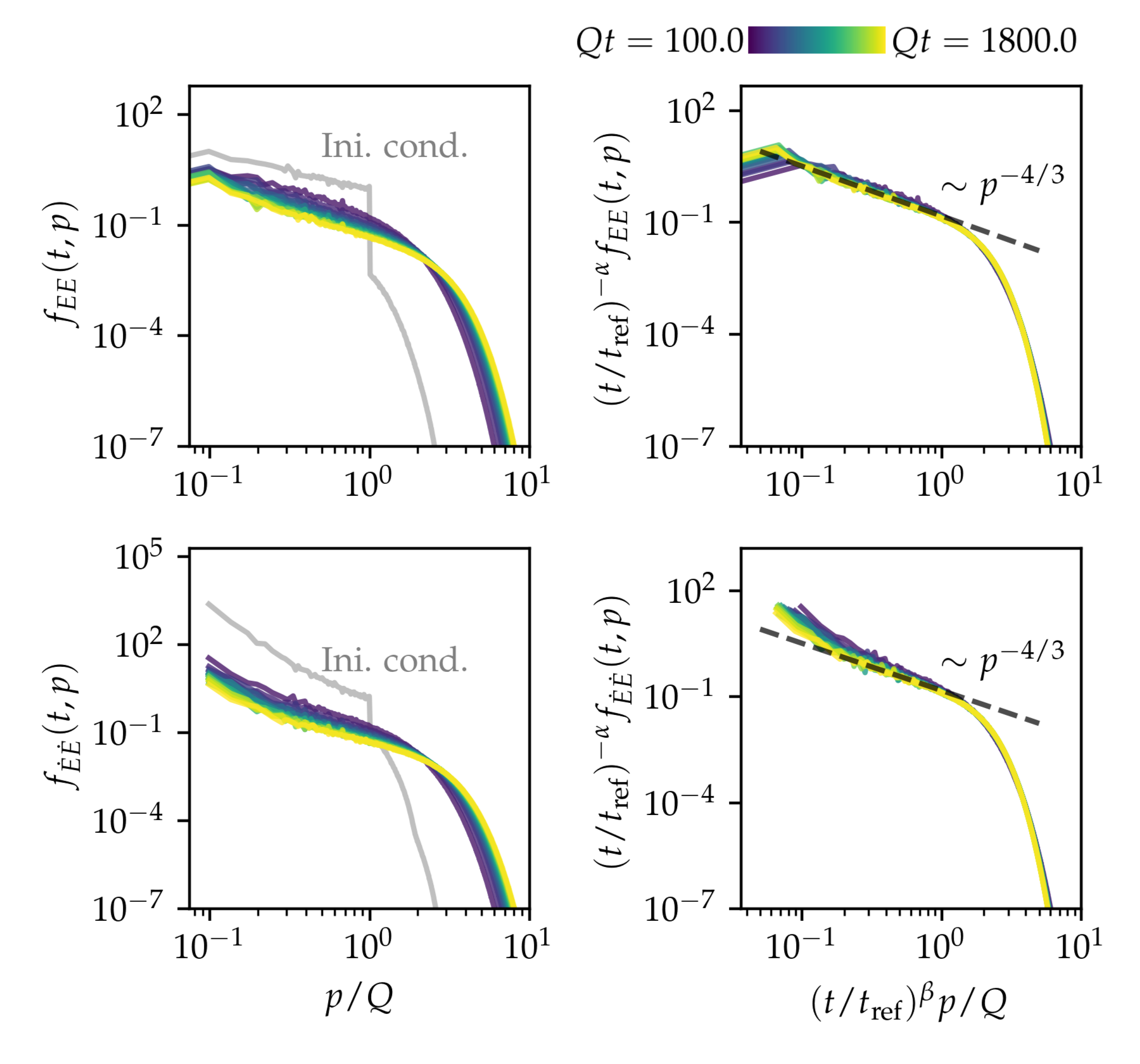}
	\caption{Occupation numbers (left) with dynamical rescaling (right) employing $\beta = -1/7$, $\alpha = 4\beta = -4/7$ in accordance with energy conservation. Top row shows occupation numbers for the $f_{EE}$ definition, bottom row for the $f_{\dot{E}\dot{E}}$ definition. Initial time displayed in gray. Dashed lines indicate $\sim p^{-4/3}$ power-laws.}\label{FigOccupations}
\end{figure}

In \Cref{FigOccupations} we display the occupation numbers for both definitions in \Cref{EqfEE_defs} across the entire time range investigated. 
Without dynamical rescaling, we observe an approximate power-law decline $\sim p^{-4/3}$ in momentum space at lower momenta, which for $f_{\dot{E}\dot{E}}$ appears overlaid by an approximate $\sim p^{-2}$ behavior of the correlator. 
Above $Q$, a steep decline towards zero appears for all times. 
We note that for both definitions occupations shift towards larger momenta in the course of time. 
For dynamical rescaling we suggestively set exponents $\beta = -1/7$ and $\alpha  = -4/7 = 4\beta$, consistent with \Cref{EqScalingAnsatzOccupations}. 
In particular, the same $\beta$ describes occupation numbers and all dynamical observables considered in the main text.

\section{The mathematics of persistent homology}\label{AppendixMathPersHom}
In this appendix we describe the construction of persistent homology groups from a more mathematical perspective. We begin with homology groups, subsequently leading to persistent homology.

\subsection{Homology groups}
Let $\mathcal{C}$ be a simplicial complex such as the alpha complexes $\alpha_r(X)$ constructed in the main text. 
The same constructions apply for cubical complexes. 
We consider chain complexes and homology groups with coefficients in $\zz_2$, such that the $k$-th chain complex $C_k(\mathcal{C})$ of $\mathcal{C}$ consists of formal sums of chains of $k$-simplices with $\zz_2$-coefficients. The boundary operator ${\partial_k:C_k(\mathcal{C})\to C_{k-1}(\mathcal{C})}$ is defined to map a chain of $k$-simplices to its boundary, which is a $(k-1)$-chain. Since boundaries of chain boundaries are empty, on has $\partial_{k-1}\circ \partial_k = 0$. We define the cycle group $Z_k(\mathcal{C}):=\ker (\partial_k)$ consisting of all closed $k$-chains, i.e., $k$-chains without boundary. We further define the boundary group $B_k(\mathcal{C}):=\mathrm{im}(\partial_{k+1})$ consisting of all $k$-chains which are boundaries of $(k+1)$-chains. We find $B_k(\mathcal{C})\subseteq Z_k(\mathcal{C})$ as subgroups, such that we can define their quotient groups,
\begin{equation}
H_k(\mathcal{C}):= Z_k(\mathcal{C})/B_k(\mathcal{C})\,,
\end{equation}
called homology groups.

We can study the topology of $\mathcal{C}$ using the homology groups $H_k(\mathcal{C})$, which capture similar topological information compared to homotopy groups of $\mathcal{C}$, but are in general not the same. Technically, elements of $H_k(\mathcal{C})$, called homology classes, are equivalence classes of $k$-cycles modulo higher-dimensional boundary contributions. We may intuitively think of homology classes as independent holes. Their number is the $\zz_2$-dimension of $H_k(\mathcal{C})$, called $k$-th Betti number,
\begin{equation}
\beta_k(\mathcal{C}) := \dim_{\zz_2}(H_k(\mathcal{C}))\,.
\end{equation}

\subsection{Persistent homology groups}
Let $\{\mathcal{C}_r\}_{r\geq 0}$ be a filtration of complexes, such as the alpha complex filtration considered in the main text. We compute all their individual homology groups $\{H_k(\mathcal{C}_r)\}_r$. In addition, the filtration contains inclusion maps $\mathcal{C}_r\hookrightarrow \mathcal{C}_s$ for all $r\leq s$. These induce maps on homology groups,
\begin{equation}
\iota_k^{r,s}: H_k(\mathcal{C}_r)\to H_k(\mathcal{C}_s)\,.
\end{equation}
The $\iota_k^{r,s}$ map a homology class in $H_k(\mathcal{C}_r)$ either to one in $H_k(\mathcal{C}_s)$, if it is still present for $\mathcal{C}_s$, or to zero, if corresponding (potentially deformed) cycles appear as boundaries in $H_k(\mathcal{C}_s)$. Further, non-trivial cokernels can appear for $\iota_k^{r,s}$: new homology classes may appear in $\mathcal{C}_s$, which are not present in $\mathcal{C}_r$. Then, $s$ can be chosen such that for sufficiently small $\epsilon> 0$,
\begin{equation}\label{EqHomologyClassBirth}
H_k(\mathcal{C}_{s-\epsilon}) \subsetneq H_k(\mathcal{C}_s)\,.
\end{equation}
We call the collection $\{(H_k(\mathcal{C}_r),\iota_k^{r,s})\}_{r\leq s}$ a persistence module. It is tame, if \Cref{EqHomologyClassBirth} holds only for finitely many distinct $s$-values. 

By the structure theorem of persistent homology \cite{Edelsbrunner2000TopologicalPersistence,ZomorodianCarlsson2004ComputingPH}, any tame persistence module is isomorphic to its persistence diagram, i.e., the collection of all its birth-death pairs $(r_b,r_d)$, $r_b<r_d\in \rr\cup\{\infty\}$. 
The same birth-death pair may appear multiple times.

\begin{figure*}
    \centering
	\includegraphics[scale = 0.77]{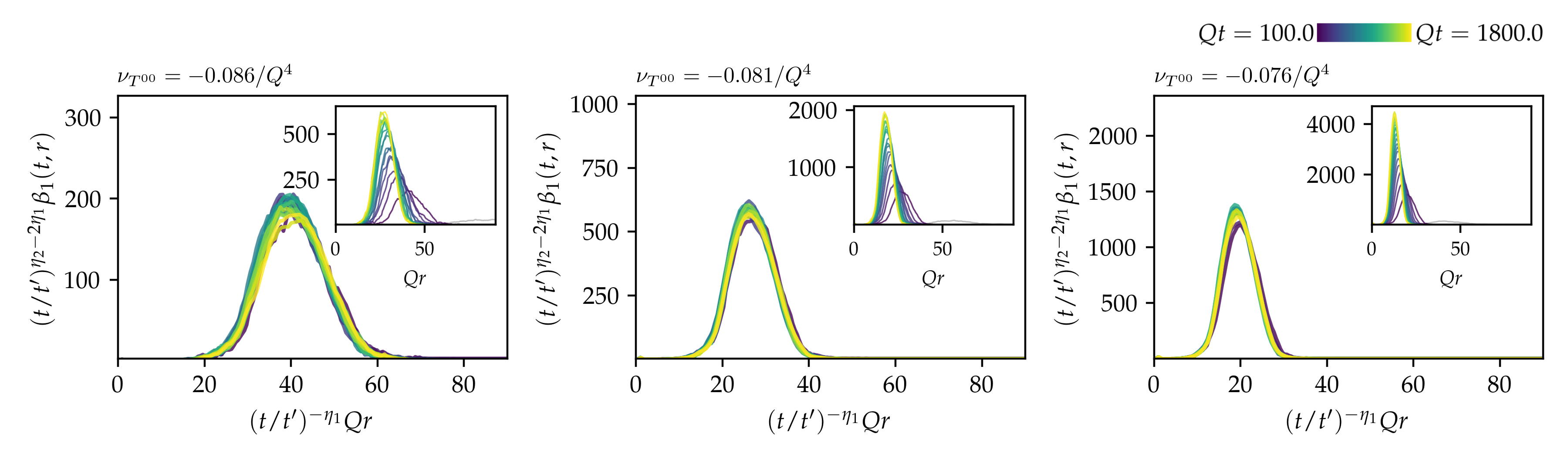}
	\caption{Dimension-1 Betti number distributions of the alpha complex filtration of $T^{00}(t,\xx)$ sublevel set for filtration parameters $\nu_{T^{00}}=-0.086/Q^4$ (left), $\nu_{T^{00}}=-0.081/Q^4$ (center) and $\nu_{T^{00}}=-0.076/Q^4$ (right). Scaling exponents are set to $\eta_1=-1/7$ and $\eta_2=5\eta_1=-5/7$ in accordance with the packing relation. Insets show figures without rescaling. Distributions at the initial time are displayed in gray.}\label{FigFiltrationParamDependenceT00}
\end{figure*}

\begin{figure}
    \centering
	\includegraphics[scale = 0.77]{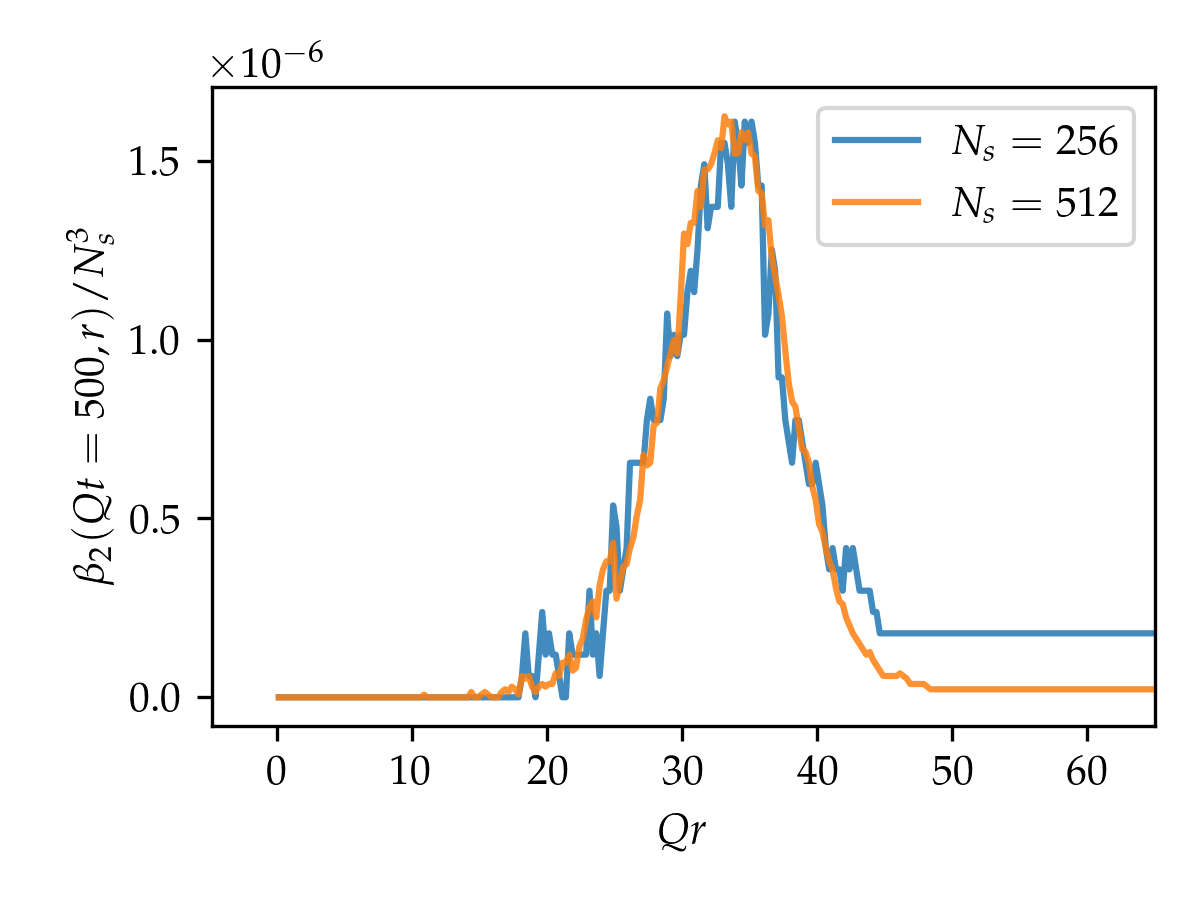}
	\caption{Dimension-2 Betti number distributions of the alpha complex filtration of $T^{00}(t=500/Q,\xx)$ sublevel sets, normalized by the volume, for filtration parameter $\nu_{T^{00}}=-0.081/Q^4$, for $N_s=256$ (blue) and $N_s=512$ (orange).}\label{FigLatticeSizeDependence}
\end{figure}

\section{Dependence on the filtration parameters and the lattice discretization}\label{AppendixFiltrationParameterDependence}
In the main text we set $\nu_{T^{00}}=-0.081/Q^4$ and $\nu_q=-0.13/Q^4$, and show results for $N_s=512$ lattice sites per spatial dimension. 
In this appendix we provide details on the dependence of the persistent homology results on these choices.
Specifically, in \Cref{FigFiltrationParamDependenceT00} we display dimension-1 Betti numbers of $T^{00}$ sublevel sets for three values of the filtration parameter $\nu_{T^{00}}$, which amount to an order of magnitude difference in the number of points in the analyzed point clouds.
While the overall magnitude of Betti numbers and the peak positions differ accordingly, up to statistical fluctuations the shape of the distributions and their time dependencies agree.
In particular, Betti number distributions can be consistently rescaled with the same pair of scaling exponents for all three $\nu_{T^{00}}$-values.
The behavior is similar for $q$ sublevel sets.

Based on mathematical theorems \cite{hiraoka2018limit,spitz2020self}, Betti numbers are expected to scale proportional to the system volume.
To verify this for our simulations, in \Cref{FigLatticeSizeDependence} we show volume-normalized dimension-2 Betti numbers for both $N_s=256$ and $N_s=512$ for the same lattice spacing.
Up to statistical fluctuations, the curves indeed agree.
Note that the different plateaus at large radii originate from the topology of the toroidal lattice itself, which is the same for all $N_s$, thus giving rise to different volume-normalized Betti numbers.

\section{Strong universality in persistent homology}\label{AppendixStrongUniversality}

\begin{figure}
    \centering
	\includegraphics[scale = 0.77]{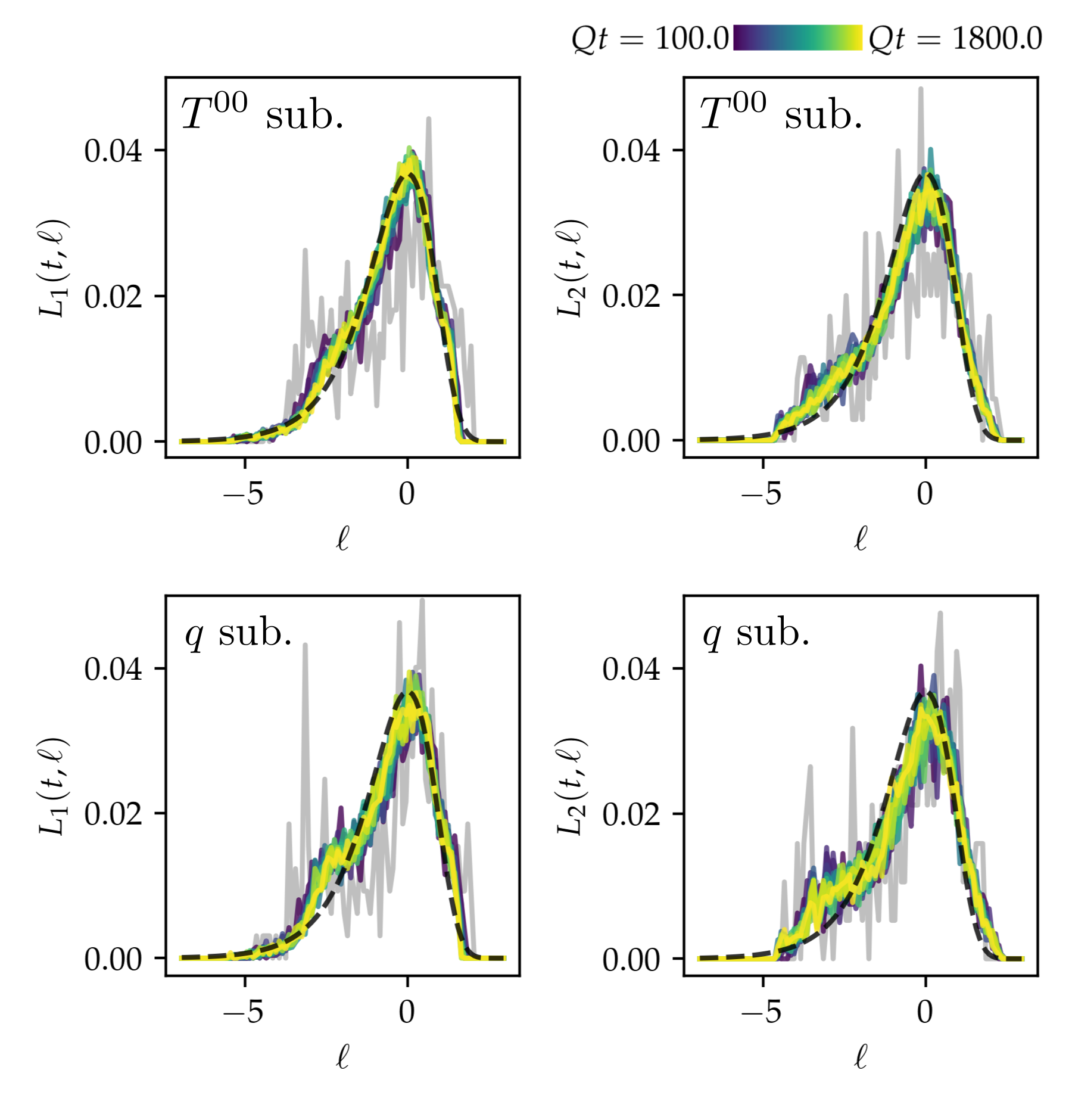}
	\caption{$\ell$-transformed distributions of persistence ratios of the alpha complex filtration for $T^{00}(t,\xx)$ (top) and $q(t,\xx)$ sublevel sets (bottom), with filtration parameters $\nu_{T^{00}} = -0.081/Q^4$ and $\nu_{q} = -0.13/Q^4$. Curves at initial time are displayed in gray. The black dashed line indicates the left-skewed Gumbel probability distribution $\sim \exp(\ell-\exp(\ell))$.}\label{FigPersistencesAlphaStrongUniversality}
\end{figure}

In this appendix we discuss transformed persistence ratios for alpha complexes of energy and topological density sublevel sets, based on the findings in \cite{bobrowski2022universality}. There, for a point cloud $X$ and the persistent homology of its alpha complexes%
\footnote{Strictly speaking, in \cite{bobrowski2022universality} the filtration of so-called \v{C}ech complexes of $X$ has been considered, which for point clouds in general position has isomorphic persistent homology to the filtration of alpha complexes.} 
it has been proposed to investigate the normalized distribution $L_k(\ell(\pi))$ of
\begin{equation}
\ell(\pi)=\frac{1}{2} \log\,\log\,\pi  - \lambda - \frac{1}{2}\bar{L}\,,
\end{equation}
where $\lambda = 0.577216...$ is the Euler-Mascheroni constant and 
\begin{equation}
\bar{L} = \frac{1}{|\dgm_k(X)|}\sum_{(r_b,r_d)\in \dgm_k(X)} \log\,\log\,\frac{r_d}{r_b}\,,
\end{equation}
$|\dgm_k(X)|$ denoting the total number of holes described by $\dgm_k(X)$.
In \Cref{FigPersistencesAlphaStrongUniversality} we show the distributions $L_k(t,\ell)$, again with the two leading lattice artefacts removed as described in \Cref{SecPersistenceResults}. We note that distributions look the same for dimensions 1 and 2. Furthermore, up to statistical fluctuations they remain constant in time, including the initial time. This provides evidence for the universality of $L_k(t,\ell)$ conjectured in \cite{bobrowski2022universality}, in which the authors suggest that the $L_k(t,\ell)$ are independent from dimensions of both the ambient space of the point clouds and the holes, and independent from the point cloud generation process, even beyond i.i.d.~processes. In particular, it is suggested that the distributions follow the left-skewed Gumbel distribution,
\begin{equation}
\label{eq:Gumbel}
L_k(t,\ell) \sim e^{\ell - e^{\ell}},
\end{equation}
also shown in \Cref{FigPersistencesAlphaStrongUniversality} as dashed curves. We find that our numerical distributions are consistent with \Cref{eq:Gumbel}. 
Remaining differences may be a consequence of sub-leading lattice artefacts that we have not removed and thus may still enter $L_k(t,\ell)$, though not visible in the persistence ratio distributions $\Pi_k(t,\pi)$ shown in \Cref{FigPersistencesAlpha}.

\begin{figure}
    \centering
	\includegraphics[scale = 0.68]{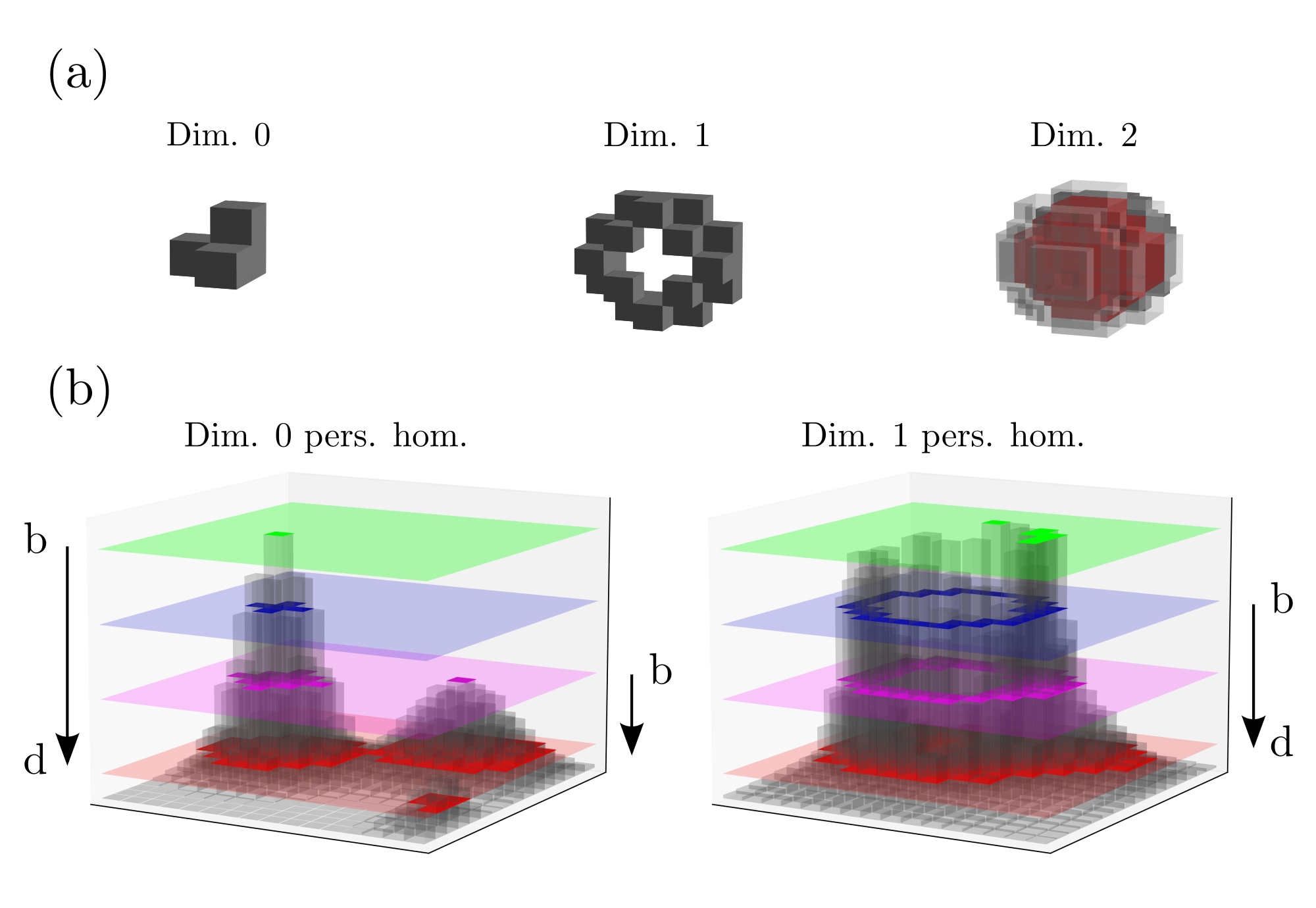}
	\caption{(a): Homology classes of different dimensions in cubical complexes, the enclosed volume is indicated in red. (b): Persistent homology classes of dimensions 0 and 1 arising from superlevel sets of a function with 2-dimensional domain. Figure adapted from \cite{Spitz:2022tul}.}\label{FigCubicalIllustration}
\end{figure}

\begin{figure*}
    \centering
	\includegraphics[scale = 0.77]{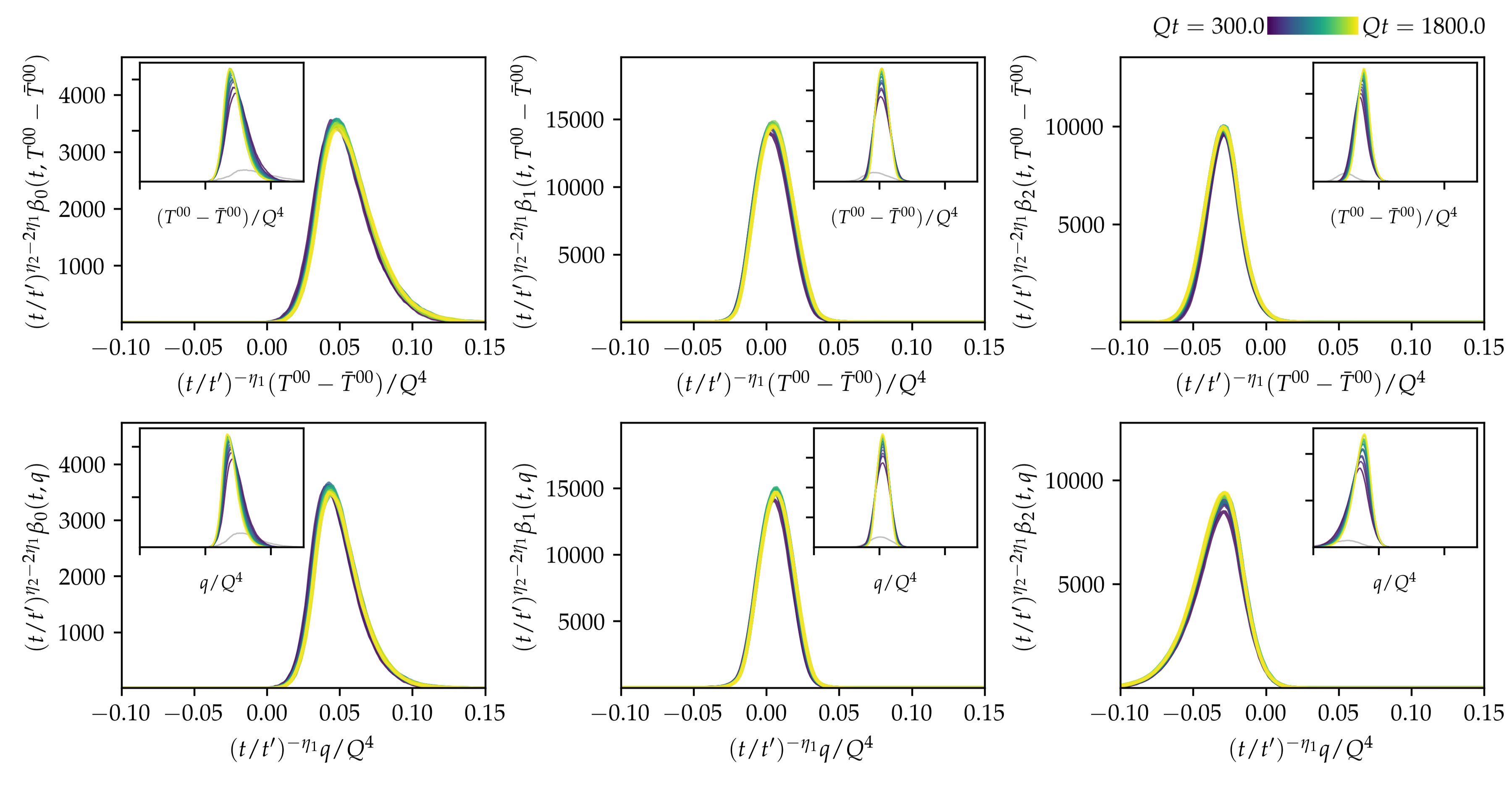}
	\caption{Betti number distributions of the $T^{00}(t,\xx)$ (top) and the $q(t,\xx)$ superlevel set filtrations (bottom) using cubical complexes. The initial time is displayed in gray. The scaling exponents are set to $\eta_1 = -1/7$ and $\eta_2 = 3\eta_1 = -3/7$. The insets show the original distributions without rescaling; the inset axes have the same ranges as the axes in the main plots.}
    \label{FigBettiCubicalSuperlevels}
\end{figure*}

\begin{figure*}
    \centering
	\includegraphics[scale = 0.77]{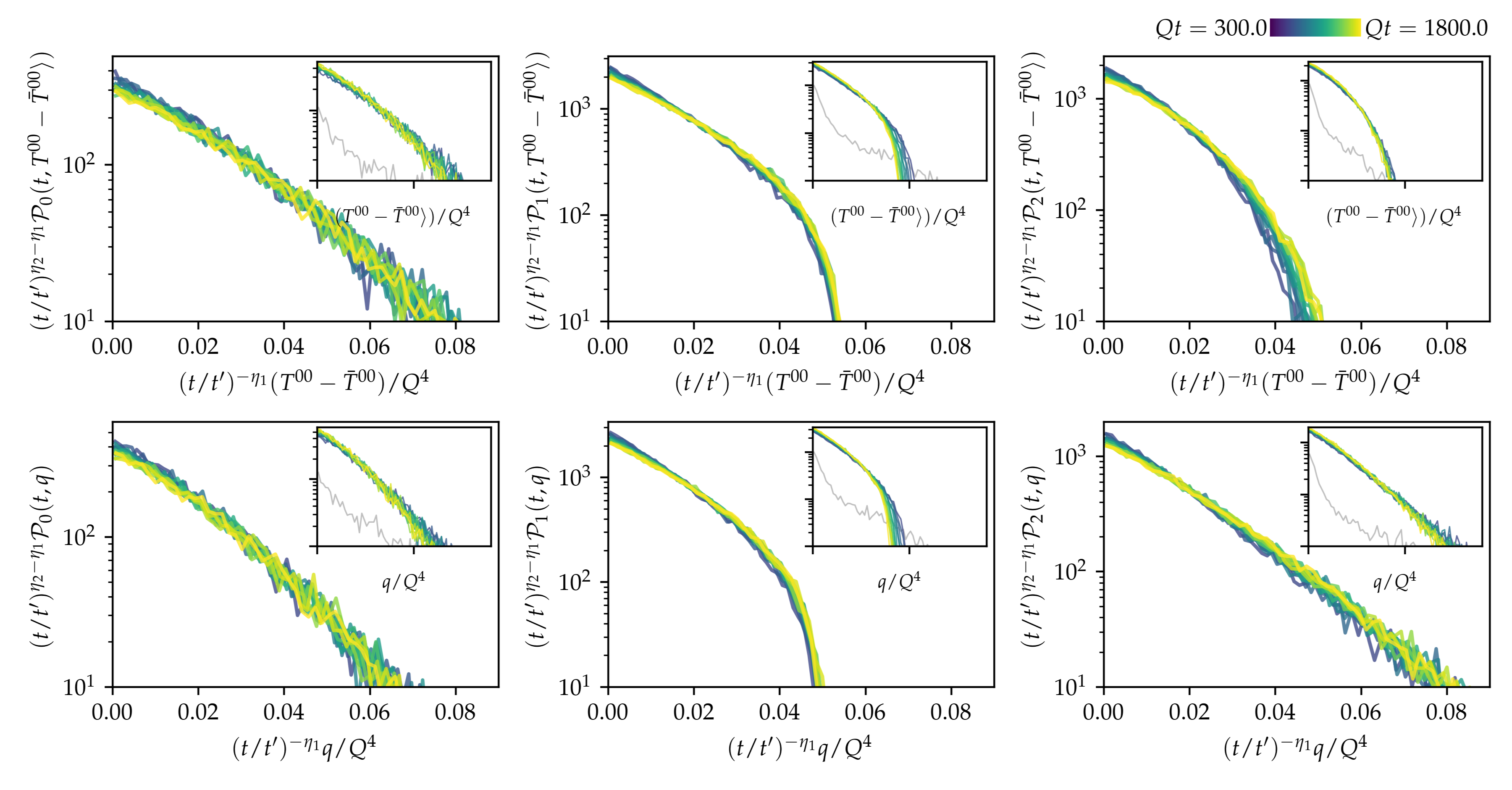}
	\caption{Persistence distributions of the $T^{00}(t,\xx)$ (top) and the $q(t,\xx)$ superlevel set filtrations (bottom) using cubical complexes. The initial time is displayed in gray. The scaling exponents are set to $\eta_1 = -1/7$ and $\eta_2 = 3\eta_1 = -3/7$. The insets show the distributions without rescaling; the inset axes have the same ranges as the main plots.}
    \label{FigPersistencesCubicalSuperlevels}
\end{figure*}

\section{Persistent homology of cubical complexes}\label{AppendixCubical}
In the main text we show results for the persistent homology of the alpha complex filtration for fixed sublevel sets, which give rise to dynamical self-similar scaling. Here, we discuss results for the filtration of superlevel set cubical complexes, with persistent homology describing persistence properties of topological structures as function values are swept through. We expect similar results for the sublevel set filtration. 

First, we introduce cubical complexes as an alternative to alpha complexes. We then discuss Betti numbers and persistence distributions, and find indications of self-similar scaling behavior. We conclude that the universal behavior stemming from the direct energy cascade also encompasses the persistent homology of cubical complexes for the superlevel set filtration.

The homology of sub- or superlevel set cubical complexes of lattice observables does not discriminate between different length scales. Therefore, lattice artefacts enter these in general. Since self-similar scaling associated to nonthermal fixed points requires a sufficient momentum space distance to lattice cutoffs, we employ spatial coarse-graining by a factor of 8 to $T^{00}_i(t,\xx)$ and $q_i(t,\xx)$ before computing the persistent homology of their superlevel set cubical complexes. Thus, higher lattice momentum scales than $\pi/(8a_s)$ do not enter the persistent homology analysis. Concerning the infrared, we have verified an approximate extensive scaling of Betti numbers and other persistent homology quantifiers with respect to the lattice volume. 

\subsection{Cubical complexes}
A cubical complex is a collection of cubes of different dimensions, which is closed under taking boundaries, similarly to simplicial complexes. We explicitly construct the filtration of cubical complexes for sub- and superlevel sets of lattice functions at a given time $t$. Cubical complexes are well-suited to describe sub- and superlevel sets of lattice functions, or pixelized data more generally.

Let $\mathfrak{C}$ be the full cubical complex of the lattice, consisting of one 3-cube $\xx + [-1/2,1/2]^3$ for each spatial lattice point $\xx\in\Lambda_s$. Such a 3-cube comes with all its faces, edges and vertices, as required to have $\mathfrak{C}$ closed under taking boundaries. On our lattice, a 3-cube is a cube of side length $a_s$, a 2-cube is a square of side length $a_s$, a 1-cube is an edge of length $a_s$, and a 0-cube is a point. In the literature, such cubes are called elementary \cite{wagner2012efficient}.

We equip $\mathfrak{C}$ with the information contained in $T^{00}_i(t,\xx)$ by inductively constructing a map $T_{t,i}:\mathfrak{C}\to \rr$. By construction of $\mathfrak{C}$, any 3-cube has a unique lattice point $\xx\in\Lambda_s$ at its center. For any 3-cube $C\in\mathfrak{C}$ we set $T_{t,i}(C):=T^{00}_i(t,\xx)$, $\xx$ the center point of $C$. Any 2-cube $D\in\mathfrak{C}$ is contained in the boundaries of two 3-cubes. For all 2-cubes $D\in\mathfrak{C}$, we set
\begin{equation}\label{EqInductiveDefTt}
T_{t,i}(D):=\min\{T_{t,i}(C)\,|\, D\subset \partial C,\, C\in\mathfrak{C}\text{ 3-cube}\}\,.
\end{equation}
Analogously, any 1-cube is contained in the boundaries of four 2-cubes, and any 0-cube is contained in the boundaries of six 1-cubes. We inductively apply \Cref{EqInductiveDefTt} to construct $T_{t,i}$ for lower-dimensional cubes from higher-dimensional ones, until $T_{t,i}$ is defined on all $\mathfrak{C}$. This construction is called the lower star filtration.

We define cubical complexes corresponding to lattice sublevel sets of $T^{00}_i(t,\xx)$ at time $t$ as
\begin{equation}
\mathfrak{C}_{T^{00}_i}(t,\nu):=T_{t,i}^{-1}(-\infty,\nu]\,.
\end{equation}
These are closed under taking boundaries, thus indeed cubical complexes. We define superlevel set cubical complexes as
\begin{equation}
\mathfrak{D}_{T^{00}_i}(t,\nu):=\mathfrak{C}_{-T^{00}_i}(t,-\nu)\,,
\end{equation}
reflecting the structure of lattice superlevel sets. For topological densities the construction is the same; simply replace $T^{00}_i(t,\xx)$ by $q_i(t,\xx)$.

We are interested in the persistent homology of the filtrations of complexes $\{\mathfrak{C}_{T^{00}_i}(t,\nu)\}_\nu$ and $\{\mathfrak{D}_{T^{00}_i}(t,\nu)\}_\nu$, as $\nu$ is swept through. Indeed, we notice that they define filtrations, e.g. $\mathfrak{C}_{T^{00}_i}(t,\nu)\subseteq \mathfrak{C}_{T^{00}_i}(t,\mu)$ whenever $\nu\leq \mu$. The previous constructions of persistent homology apply also in this setting. The persistent homology of cubical complex filtrations can be efficiently calculated \cite{wagner2012efficient}. We again use GUDHI \cite{10.1007/978-3-662-44199-2_28} to compute persistent homology of the cubical complex filtration, and use periodic cubical complexes to take spatially periodic boundary conditions into account.

In \Cref{FigCubicalIllustration}(a) we illustrate the meaning of cubical complex homology classes. In \Cref{FigCubicalIllustration}(b) we sketch the meaning of birth $b$ and death $d$ of homology classes as the filtration of superlevel sets is swept through. We note that the birth of dimension-0 homology classes happens at function values of local maxima. Their persistences quantify their dominances, dying when merging at saddle points with tails of other local maxima. Dimension-1 homology classes can be thought of as describing potentially deformed volcano-like features in function landscapes, and analogously for higher-dimensional homology classes.

\subsection{Betti number distributions}

In \Cref{FigBettiCubicalSuperlevels} we show results for Betti number distributions of all dimensions for both the energy and the topological density superlevel set filtrations using cubical complexes. 
Insets show numbers without, and main plots with dynamical rescaling. Note that energy densities are shown with the average subtracted. We notice from the unrescaled figures that topological features of both energy and topological densities give rise to a hierarchy of peaks between dimensions. For superlevel sets of a function whose values scatter approximately symmetrically around zero, connected components appear first at positive filtration parameters. 
Multiple such connected components need to merge to form a dimension-1 feature, which is born around zero. 
Typically at negative filtration parameters, multiple dimension-1 features die to give birth to a dimension-2 feature.

In the course of time, throughout dimensions and energy and topological densities topological features shift towards zero filtration parameters suggesting a homogenization of structures. Further, their number increases, indicative of their dynamical refinement. In the main plots we suggestively rescaled the figures with $\eta_1 = -1/7$ and $\eta_2 = -3/7 = 3\eta_1$. We notice approximate matching of the rescaled curves. Notice that aside of the initial condition we here show data from $Qt=300$ onwards, while in the main text we show data starting with $Qt=100$. Betti number distributions do not allow sufficiently consistent rescaling before $Qt = 300$ (not displayed). 

The exponent $\eta_1 = -1/7$ is consistent with the earlier findings of $\eta_1$ for alpha complexes and $\beta$ for correlation functions. However, here it describes the dynamical refinement of structures in the space of function values, not with respect to spatial sizes of structures. How can the exponents still agree? We recall from \Cref{SecCorrEnergyTopDensities} that the correlators $\langle (T^{00}(t,p_x))^2\rangle_c$ and $\langle (q(t,p_x))^2\rangle_c$ scale overall as $\sim t^\beta$ for spatial scaling exponent $\beta = -1/7$. If such correlators contribute predominantly to local fluctuations in $T^{00}_i(t,\xx)$ and $q_i(t,\xx)$, we can heuristically find that topological features scale in the space of function values as $\sim t^\beta$, too. We take our observation of $\eta_1 = -1/7$ for the superlevel set filtration as an indication for this.

The number of topological features which appear in $T^{00}_i$ superlevel sets is bounded by total energy conservation in a similar way that the number of topological features for alpha complexes is bounded by the constant system volume. This effectively provides a one-dimensional constraint for the persistent homology of the superlevel set filtration. Then, for cubical complexes of $T^{00}_i$ superlevel sets the packing relation yields $\eta_2 = 3\eta_1$ \cite{Spitz:2020wej, spitz2020self}, consistent with the data.

The apparent similarity of energy and topological density cubical complexes in persistent homology can again be understood from independent spatial and colour directions of $\EE(t,\xx)$ and $\BB(t,\xx)$ as proposed in \Cref{SecCorrEnergyTopDensities}.
Yet, differences are visible in particular for the left-sided tail of dimension-2 Betti numbers, where the Betti number distribution has larger support for topological rather than energy densities.

\subsection{Persistence distributions}
In \Cref{FigPersistencesCubicalSuperlevels} we display the distribution of absolute persistences, $d-b$, for energy and topological density superlevel sets. Again, insets show figures without rescaling, which we discuss first. We notice that dimension-0 distributions decline approximately exponentially with persistence values. Dimension-1 distributions have more restricted support only up to $T^{00}-\bar{T}^{00},q\simeq 0.05\, Q^4$. Persistence distributions of dimension-2 features of energy densities have more restricted support compared to topological densities, whose dimension-2 persistence distributions moreover show approximate exponential behavior. This indicates that energy density persistences have an upper bound due to the positivity of $T^{00}_i$, which is not the case for $q_i$. There, the approximate symmetry among dimension-0 and dimension-2 features hints at a symmetric distribution of topological density values around zero, which in particular features local lumps, similarly to the deductions in \cite{Spitz:2022tul}.

Let us now consider absolute persistences defined as integrals of the averaged dimension-$k$ persistence pair distributions over the death parameter,
\begin{align}
    \mathcal{P}_k(t,p) = \int_0^\infty \dd d\, \langle \Pfrak_k\rangle(t,d - p, d)\,.
\end{align}
The scaling ansatz \Cref{EqPersPairDistribScalingAnsatz} then leads to the self-similar scaling
\begin{align}
\mathcal{P}_k(t,p) = (t/t')^{\eta_1-\eta_2} \mathcal{P}_k(t',(t/t')^{-\eta_1}p)\,.
\end{align}
Accordingly, we have rescaled the curves in the main plots in \Cref{FigPersistencesCubicalSuperlevels} using the same exponents $\eta_1 = -1/7$ and $\eta_2 = -3/7$ as for the Betti numbers shown in \Cref{FigBettiCubicalSuperlevels}. We observe approximate agreement among the rescaled curves mostly up to statistical fluctuations, which is indicative of dynamical self-similarity. We suggest that the deviations for dimension-2 persistences of energy density sublevel sets are due to the positivity bound.

\bibliography{literature}

\end{document}